\documentclass{aa}  

\usepackage{graphicx}
\usepackage{subcaption}
\usepackage{booktabs}
\usepackage{tabularx}
\usepackage{amsmath}
\usepackage{xcolor}
\usepackage[utf8]{inputenc}
\usepackage{textgreek}
\usepackage{comment}
\usepackage{txfonts}
\usepackage{threeparttable}
\usepackage[]{hyperref}
\hypersetup{colorlinks,citecolor=blue, linkcolor=blue, urlcolor=blue, breaklinks=true}

\usepackage[normalem]{ulem}

\def\msun{\rm\, {M_\odot}}
\def\kms{\rm\, km\,s^{-1}}

\def\ramses{\textsc{ramses}}
\def\dynesty{\textsc{dynesty}}
\def\barolo{\textsc{$^\mathrm{{3D}}$Barolo}}
\def\galpynamics{\textsc{galpynamics}}
\def\galfit{\textsc{galfit}}

\begin{document}

   \title{The galaxy-halo connection and the dynamical evolution of a giant disc in a massive node of the Cosmic Web at $z\sim3$}
   
   \titlerunning{The galaxy-halo connection and dynamical evolution of a giant disc galaxy at $z\sim3$}

   \author{G. Quadri\inst{1}
        \fnmsep\thanks{g.quadri2@campus.unimib.it} 
          \and
          S. Cantalupo \inst{1}
          \and
          C. Bacchini \inst{2}
          \and
          A. Pensabene \inst{1,3,4}
          \and
          A. Lupi \inst{5,6,7}
          \and
          G. Pezzulli \inst{8}
          \and
          W. Wang \inst{1}
          \and
          M. Galbiati \inst{1}
          \and \\
          T. Lazeyras \inst{1}
          \and
          N. Ledos \inst{1}
          \and
          H. Mao \inst{9, 1}
          \and
          A. Travascio \inst{10}
          }

   \institute{
            Dipartimento di Fisica G. Occhialini, Universit\`a degli Studi di Milano-Bicocca, Piazza della Scienza 3, I-20126 Milano, Italy
            \and
            DARK, Niels Bohr Institute, University of Copenhagen, Jagtvej 155, 2200 Copenhagen, Denmark
            \and
            Cosmic Dawn Center (DAWN), Copenhagen, Denmark
            \and
            DTU Space, Technical University of Denmark, DK-2800 Kgs. Lyngby, Denmark
            \and 
            Como Lake Center for Astrophysics, DiSAT, Universit\`a degli Studi dell'Insubria,  via Valleggio 11, I-22100, Como, Italy
            \and
            INFN, Sezione di Milano-Bicocca, Piazza della Scienza 3, I-20126 Milano, Italy
            \and 
            INAF, Osservatorio Astronomico di Bologna, Via Gobetti 93/3, I-40129 Bologna, Italy
            \and
            Kapteyn Astronomical Institute, University of Groningen, NL-9747 AD Groningen, the Netherlands
            \and
            Purple Mountain Observatory, Chinese Academy of Sciences, Nanjing 210023, China
            \and 
            INAF – Osservatorio Astronomico di Trieste, I-34131 Trieste, Italy}

   \date{Received \today; accepted}

 
  \abstract{
  Recent JWST observations revealed the surprising presence of a giant and massive disc galaxy in a Cosmic Web node at z$\sim3$. This galaxy, named the Big Wheel, has a size almost three times larger than expected for typical disc galaxies at the same redshift and similar stellar masses. Constraining the origin and formation history of the Big Wheel requires knowledge of its dark matter halo properties, which are difficult to derive from JWST observations alone. Here, we investigate the dark matter halo of the Big Wheel and provide further constraints on the galaxy baryonic content, combining a physically motivated dynamical model with deep ALMA kinematical data. By using priors based on JWST photometric data and CO kinematics, we infer a dark matter halo mass of $\log (M_{h}/\msun)= 12.11^{+0.29}_{-0.17}$ and a stellar mass of $\log(M_{\star}/\msun)=11.00^{+0.11}_{-0.12}$, resulting into a stellar-to-halo mass ratio of $M_\star/M_h=0.06^{+0.04}_{-0.03}$. This value is significantly higher than expected from empirical stellar-to-halo-mass relations currently available in the literature. This implies that the Big Wheel may have assembled its stellar content in a much more efficient way with respect to the general galaxy population at this redshift. Combined with its morphological properties, our results suggest that the Big Wheel had a tranquil recent formation history, with probably no major mergers, violent disc instabilities, or strong ejective feedback. We perform a numerical simulation of an idealised galaxy and let it evolve adiabatically for $2.5$ Gyr to demonstrate that the modelled galaxy does not develop gravitational instabilities during its evolution that could alter its resemblance to its currently observed state. Although systems alike the Big Wheel are arguably rare, our results offer new constraints on the contribution of accretion and feedback to the formation history of the most massive discs within high-redshift Cosmic Web nodes.
  }


   \maketitle
%

\section{Introduction}
According to our current cosmological paradigm, galaxy formation and evolution occur within dark matter (DM) halos, virialised and gravitationally self-bound structures that form hierarchically \citep{White78}. Within this framework, the properties of the baryonic matter contained in galaxies, such as their stellar masses, sizes and angular momentum, should be closely linked to their DM halos, as found in several observational studies \citep[see e.g.,][]{Wechsler18, Paquereau25}.

One way to study such galaxy-halo connection is by exploring the relationship between the mass of the DM halo and the stellar mass of the galaxy that it hosts, usually referred to as the stellar-to-halo mass relation (SHMR). This quantity encodes information on the relative efficiency of galaxy stellar and halo mass build-up. While the DM halo mass growth depends only on gravitational mechanisms, the stellar mass build-up depends instead on a series of so far poorly constrained processes, including baryonic accretion and feedback \citep{FaucherGiguere11, Hopkins14, Somerville15}. As such, the SHMR as a function of the halo mass could be effectively used to constrain the role of these mechanisms in galaxy formation and evolution. By combining empirical and observational constraints, e.g. based on halo abundance matching \citep{Moster13, Moster18, Behroozi13, Girelli20, Shuntov22}, several studies suggest that the SHMR as a function of the halo mass for a typical galaxy (independent of its morphological type or other properties) has a peak value at characteristic halo mass that significantly changes according to the considered redshift.
At lower and higher halo masses, the function decreases very rapidly. This is usually interpreted as an effect of galaxy ejective or preventive feedback. For instance, lower values of the DM halo masses imply shallower gravitational potentials. For these systems, the combined effect of stellar feedback mechanisms – such as stellar winds, supernova explosions (SNe), photoheating, and radiation pressure – is believed to be efficient enough to heat and expel the gas from the galaxy and the halo \citep{Dekel86, MacLow99, Somerville15}. 
On the contrary, for high-mass halos, the SHMR decline is commonly attributed to the onset of (super)massive black-hole feedback given the need of a much higher quantity of energy to overcome the deeper potential wells \citep{Croton06, silk12, Behroozi19}. 
At $z\sim 0$, the bending of the SHMR is primarily driven by the dominance of early-type galaxies at the high-mass end, while spirals exhibit a SHMR that scales linearly with mass \citep{Posti21}. However, the extension of this trend to $z>0$ has been hampered because of observational limitations, such as poor signal-to-noise ratios (SNR) and restricted spatial resolution.

Since both the DM halo and stellar masses of individual galaxies always grow with time, the value of the SHMR for a particular system encodes information on its overall formation history, including accretion, star-formation, and feedback efficiency. However, the aforementioned empirical models provide, by construction, a representation of the SHMR for a statistical sample and not for particular, individual galaxies for which such information would be equally vital to understand their formation histories. For the latter, a direct constraint of their DM halo mass is needed.

For instance, recent studies have revealed the presence of galaxies with peculiar physical properties within highly overdense regions, such as the Big Wheel — a giant disc galaxy located within the MQN01 Cosmic Web node at $z\approx3.2$ \citep[][W25 hereafter]{Wang25}. This galaxy has a stellar mass of $\log(M_\star/\msun) = 11.37 \pm 0.20$ and a half-mass radius equal to $6.3^{+0.7}_{-0.6}$ kpc, which is at least around $3$ times larger than expected for typical galaxies with similar masses at a comparable redshift \citep{vanderWel14, Ormerod24} and surprisingly similar to present-day super-spiral galaxies \citep{DiTeodoro21, DiTeodoro23}. What is the origin of such a large disc at such early cosmic epochs? One possibility is that the Big Wheel would reside in a very large and massive DM halo, much larger than expected for its stellar mass, according to the known relation between DM halo and galaxy disc sizes, which indicates that the size of a galaxy is approximately $2\%$ of the virial radius \citep{Kravtsov13, Somerville18}. Alternatively, in the framework of the theoretical models of \cite{Mo98}, the large disc size of the Big Wheel could be attributed to the large spin parameter $\lambda$ of its DM halo. In order to disentangle these scenarios, knowledge of the Big Wheel DM halo mass is paramount. At the same time, the value of the stellar-to-halo mass (SHM) ratio of the Big Wheel could provide a powerful framework to connect its stellar content with the mass of its DM halo, offering insights into the role of accretion and feedback in its formation history. One approach is to perform a detailed mass decomposition of galaxy rotation curves, separating the contributions from stars, gas, and DM. However, directly constraining the DM distribution remains challenging due to the so called disc-halo degeneracy \citep{vanAlbada85}, according to which the observed rotation curve can be equally well reproduced by different combinations of disc and halo masses, making it difficult to uniquely determine the contribution of each component.

In this work, we make use of dynamical mass decomposition techniques to derive physically robust estimates of the DM halo mass for the Big Wheel, providing a measurement of the SHM ratio for the system. Furthermore, we simulate an idealised galaxy imposing as initial conditions the results based on the dynamical model for the Big Wheel, and we check for the gravitational stability of the system, letting it evolve for $2.5$ Gyr — up to $\approx8$ times its dynamical timescale.

This work is organized as follows: in Section~\ref{sec:overview_data}, we briefly summarise the available observational data and measurements on the Big Wheel. In Section~\ref{sec:dyn_est}, we present the dynamical modelling adopted to infer the DM halo properties and describe the corresponding results. In Section~\ref{sec:num_sim}, we illustrate the numerical simulation performed and present its main outcomes. The implications of our findings are deeply discussed in Section~\ref{sec:discussion}. Finally, in Section~\ref{sec:conclusion}, we draw our conclusions.

Throughout this paper, we adopt a standard $\Lambda$CDM cosmology with $H_0 = 67.66 \ \mathrm{km \ s^{-1} \ Mpc^{-1}}$, $\Omega_{\mathrm{m}}= 0.3111$, and $\Omega_{\mathrm{\Lambda}} = 1 - \Omega_{\mathrm{m}}$ from \citet{Planck20},

\section{Observational datasets}
\label{sec:overview_data}
\subsection{Rest-frame UV and optical}
\label{sec:optical_data}
Our current knowledge of properties of the baryonic component of the Big Wheel galaxy relies on a deep multi-wavelength dataset obtained in the rest-frame UV and optical (W25 and \citealt{Galbiati25}), in addition to ALMA (\citealt{Pensabene24} and Pensabene et al., in prep.). In particular, JWST observations at rest-frame wavelengths of 0.4 $\rm \mu m$ to 0.8 $\rm \mu m$ 
revealed that the Big Wheel is a giant disc galaxy at $z \approx 3.25$ with an half-light radius of $9.6^{+0.5}_{-1.2}$ kpc measured at rest-frame $0.5 \mathrm{\mu m}$, making it $3$ times larger than expected for typical star-forming galaxies at similar stellar mass and redshift (W25). At a rest-frame wavelength of 0.2 $\rm \mu m$ in MUSE and HST observations, the Big Wheel appears as a collection of isolated clumps due to moderate dust obscuration. The galaxy velocity maps measured from the ALMA CO(4--3) and the H$\rm \alpha$ emission lines derived from JWST/NIRSpec confirmed the rotating-disc nature of the Big Wheel. The Big Wheel is not found in isolation but located within the MQN01 structure, hosting one of the highest overdensity of galaxies and active galactic nuclei (AGNs) discovered so far at this redshift \citep{Pensabene24, Pensabene25, Galbiati25, Travascio25b, Travascio25}. Moreover, Chandra X-ray observations of the Big Wheel reveal the presence of an AGN at the centre of the galaxy, a result independently supported by nebular emission-line ratios in the central NIRSpec spectrum.

Here, we briefly describe the properties of the galaxy derived from the observations above. In particular, we present the range of values for the total stellar mass, half-mass radius and bulge-to-disc ratio used as a prior for our Bayesian analysis and how they were obtained. The constraints on the distribution of baryons are essential to break the degeneracy with the DM halo. The stellar mass prior is adopted from the values derived by \citet{Galbiati25}, i.e. $\log(M_{\star}/\msun) = 11.37 \pm 0.20$, obtained by fitting the Spectral Energy Distribution (SED) of the Big Wheel with the CIGALE code (\citealt{Burganella05, Noll09, Boquien19}, version 2022.1, which includes X-Ray implementation by \citealt{Yang20, Yang22}).They assumed \citet{Bruzual03} stellar population synthesis models, a \citet{Chabrier03} initial mass function and the dust attenuation law of \citet{Calzetti00}, together with an exponentially declining star formation history. The quoted uncertainty of $0.20$ dex accounts for both the photometric uncertainties and the dominant contribution from systematic effects associated with the choice of the stellar population models, as described in \cite{Galbiati25}. 

The half-mass radius has been derived following the procedure described in the Supplementary Figure 1 in W25, i.e. performing SED fitting within radial bins considering all available photometric bands. Through this procedure, we obtained $r_{\rm half-mass} =6.3 \rm \ kpc$, which is slightly different from the value published in W25 ($r_{\rm half-mass} =6.7 \rm \ kpc$) due to the finer radial binning adopted in this work See Appendix~\ref{appendix:SED} for more details. From the same procedure we obtained a bulge-to-total mass fraction of $B/T = 0.13 \pm0.05$. Finally, the inclination and the position angle of the galaxy are fixed to the values of $i=38°$ and P.A.$=34°$, respectively, obtained by combining morphological and kinematic constraints (see Appendix~\ref{appendix:kinematic_mod} for more details). We note that, in the present spatially resolved SED analysis, the dust continuum emission from the high resolution ALMA observations (Pensabene et al., in prep.) is not taken into account. Consequently, the inferred physical parameters may change once this is included.

\subsection{ALMA CO data}
\label{sec:almaC0}

\subsubsection{Low angular resolution dataset}
\label{sec:lowres_data}
In this work, we benefit from ALMA Cycle 8 data acquired in band 3 (Program ID: 2021.1.00793.S, see \citealt{Pensabene24} for full details). The Big Wheel galaxy has been identified in low-angular resolution ($\sim1.3{\arcsec}$) ALMA mosaic observations covering $\sim 4\,{\rm arcmin^2}$ around the MQN01 structure ($z\approx3.25$), and detected in both the (observed-frame) 3-mm dust continuum and its CO(4--3) line emission. For this work, we reprocessed the data to obtain a "cleaned" ALMA datacube around the Big Wheel galaxy by imaging the raw visibilities with the {\tt tclean} task of the Common Astronomy Software Application (CASA; \citealt{McMullin07}). We set the phase centre at the location of the galaxy (ICRS 00:41:35.129 -49:37:12.402), we adopted {\tt "briggs"} as weighting scheme of the visibilities with a robust parameter of $R=0.5$ to optimize between the surface brightness sensitivity and the angular resolution. We chose a pixel size of $0\rlap{.}{\arcsec}1$ to achieve the Nyquist sampling of the longest baselines and a channel width of $25\,{\rm km\,s^{-1}}$, and we cleaned down to $1.5\times{\rm RMS}$ per channel within a circular mask of $2\arcsec$ radius encompassing the whole line emission of the galaxy. The cleaned datacube has a briggs-weighted synthesized beam size of $1\rlap{.}{\arcsec}17\times1\rlap{.}{\arcsec}05$ at the reference frequency of $108.6\,{\rm GHz}$, and a noise RMS of $\approx0.3\,{\rm mJy\,beam^{-1}\,channel^{-1}}$. We then modelled the continuum in the line-free channels with a constant and subtract it from the datacube by using the CASA {\tt imcontsub} task. In this work, we use this final continuum-subtracted datacube to model the observed cold gas kinematics as traced by the CO(4--3) line (see Appendix~\ref{sec:gaskin}).

\subsubsection{High angular resolution dataset}
\label{sec:highres_data}

We followed-up the Big Wheel galaxy with ALMA in Cycle 12 (Program ID: 2025.1.00107, PI: Pensabene) targeting the CO(4--3) line emission at higher resolution ($\sim 0\rlap{.}{\arcsec}35$) employing $\approx27$ hours of total telescope time in C-6 array configuration. The full details of the reduction, processing and analysis of this dataset, as well as a comprehensive study of the cold gas kinematics and morphology will be presented in Pensabene et al. (in prep.). In this work, we benefit from this higher-quality and deeper data to improve both the sampling and the extension of the molecular gas rotation curve of the galaxy (see, Section~\ref{sec:results_dynmod}). However, we emphasize that the dynamical model results obtained from the low angular resolution dataset are fully consistent with those derived from the high angular resolution dataset. 

\section{Dynamical modelling}
\label{sec:dyn_est}
In this section, we describe the dynamical model used to infer the masses of the baryonic and DM components of the Big Wheel.
In particular, by comparing the model to ALMA CO kinematical data described in Section~\ref{sec:almaC0}, we aim to obtain the Big Wheel DM halo mass and to provide additional constraints on its baryonic components. In particular, the CO(4--3) emission line accurately traces the cold gas distribution and kinematics, offering a reliable proxy for the circular velocity curve of the galaxy.

\subsection{Rotation curve decomposition}
\label{sec:dyn_mod}
In order to model the circular velocity of the Big Wheel, we followed the same approach as described in \cite{RomanOliveira24, RomanOliveira26}. Briefly, we assumed the circular velocity of the galaxy as:
\begin{equation}
\label{eq:circvel}
    V_{\rm c, tot}^{2} = \sum_{i=1}^{n}R\left( \frac{\partial\Phi_{i}}{\partial R}\right) = \sum_{i=1}^{n} V_{\mathrm{c}, i}^{2}\text{,}
\end{equation}
where $R$ is the radius, $\Phi_{i}$ is the gravitational potential, and $V_{\mathrm{c}, i}$ is the circular velocity of the $i$-th mass component. In our model, the total potential is decomposed into the contributions from the DM halo, the stellar and gaseous discs, and the stellar bulge. Then, we fitted the Big Wheel rotation curve derived from the ALMA CO(4--3) emission line, using Eq. (\ref{eq:circvel}). To infer the posterior distribution of key parameters, we used the nested-sampling algorithm \textsc{dynesty} \citep{Speagle20}. In the following, we provide more details on the choice of priors and on the main model assumptions component by component. Table~\ref{tab:priors} provides a summary of the prior distributions and their associated ranges.

\subsubsection{Dark matter halo}
\label{sec:dm}
We assume that the density of the DM halo follows a spherical Navarro–Frenk–White (NFW) density profile \citep{Navarro96}, given by
\begin{equation}
    \rho_{\mathrm{DM}}(r)=\rho_{\mathrm{NFW}}(r) \left(1-\frac{\Omega_{b}}{\Omega_m}\right)= \frac{\rho_{\mathrm{crit}} \ \delta_{\mathrm{c}}}{(r/r_{\mathrm{s}})(1+r/r_{\mathrm{s}})^2}\left(1-\frac{\Omega_{b}}{\Omega_m}\right)\text{,}
\end{equation}
where $\rho_{\mathrm{crit}}=3H_{0}/8\pi G$ is the critical density of the Universe, $\Omega_b$ and $\Omega_m$ represent the density parameters for baryons and total matter, respectively, expressed as fractions of the critical density of the Universe. The scale radius is defined as $r_{\mathrm{s}} = r_{200}/c$, with $c$ being the halo concentration and $r_{200}$ the radius enclosing a mean density 200 times the critical one. From this latter definition, the characteristic overdensity of the halo $\delta_{\mathrm{c}}$ is related to $c$ through
\begin{equation}
    \delta_{\mathrm{c}}=\frac{200}{3} \frac{c^3}{[\mathrm{ln}(1+c) -c/(1+c)]}\text{.}
\end{equation}
The concentration varies with the virial mass through the concentration-mass relation of \citet{Dutton14}
\begin{equation}
\label{eq:concentration}
    \mathrm{log}_{10}\ c=a+b\ \mathrm{log}_{10} (M_{200}/[{10^{12} \ h^{-1} \ \msun}]),
\end{equation} 
where $h^{-1}$ is the reduced Hubble constant, $a$ and $b$ change according to $z$ as $a=0.520+(0.905-0.520)\exp(-0.617 \ z^{1.21})$, $b=-0.101 + 0.026 \ z$.

We define the prior on $M_{\rm DM}$ as $M_{\rm DM} ={(M_{\star} + M_{\mathrm{gas}})}/{f_{\mathrm{bar}}}$ where $f_{\mathrm{bar}}$ is the baryonic fraction. In particular, $f_{\mathrm{bar}}$ is sampled from a logarithmic uniform prior, with values ranging between $10^{-4}$, which corresponds to a heavily DM dominated galaxy, and the cosmological baryon fraction, i.e. $f_{\rm bar, c} = \Omega_{b} / (\Omega_{m}-\Omega_{b}) \approx 0.187$ from \cite{Planck20}.
The concentration $c$ was drawn from a lognormal distribution centred on the mean value predicted by Eq. (\ref{eq:concentration}) and with an intrinsic scatter of $0.11$ dex, consistent with results from \citet{Maccio}. We note that lognormal distributions are chosen in order to ensure that also small values are well sampled. The corresponding contribution of the dark matter to the virial velocity — representing the circular velocity at the virial radius $r_{200}$ — is computed as
\begin{equation}
    V_{\mathrm{DM}} = \left(\frac{G M_{\rm DM} H(z)}{100}\right)^{1/3}\text{.}
\end{equation}

\subsubsection{Gaseous and stellar discs}
\label{sec:discs}
The disc components are modelled assuming an exponential profile:
\begin{equation}
\label{eq:density_discs}
    \rho(R, h) = \Sigma_0 \ \mathrm{exp}\left(-\frac{R}{R_{d}}\right) \times f(h) \textit{,}
\end{equation}
where $R_{d}$ is the scale radius and $\Sigma_0$ is the central surface density. The term $f(h)$ is a function of the height $h$ above the galaxy equatorial plane ($h = 0$) and describes the mass vertical distribution. For both the gaseous and the stellar disc, we consider a thick disc with a hyperbolic secant squared vertical profile
\begin{equation}
    f(h) = \frac{1}{2h_\mathrm{d}} sech^2 \left(\frac{h}{h_\mathrm{d}}\right) \text{,}
\end{equation}
with $h_\mathrm{d}=R_{d}/7$ \citep{2026vanAsselt}. The circular velocities of the thick exponential discs are computed using the \galpynamics \footnote{https://gitlab.com/iogiul/galpynamics} package, where the gravitational potential is obtained by numerically integrating the full three–dimensional density distribution in Eq. (\ref{eq:density_discs}) over the radial and vertical coordinates. The resulting potential is differentiated with respect to the radial coordinate to obtain the circular velocity in the equatorial plane.

We adopt a lognormal prior for the stellar mass using a distribution centred on the fiducial value reported in \citet{Galbiati25} and in Section~\ref{sec:overview_data}. The standard deviation of the distribution is assumed to be $0.2$ dex. Similarly, we considered a lognormal prior for the radii of the stellar and gaseous discs from the half-mass radius reported in Section~\ref{sec:overview_data}, according to the relation $R_{d, \rm disc}=r_{\rm half}/1.678$, valid for exponential discs.

The prior on the gaseous disc mass is derived starting from the CO(4--3) line luminosity $L'_{\mathrm{CO(4-3)}}$ as  $M_{\rm gas} = \beta L'_{\mathrm{CO(1-0)}} = (\alpha_{\rm CO}/ r_{41})L'_{\mathrm{CO(4-3)}}$, where $\alpha_{\rm CO}$ is the luminosity-to-molecular gas mass conversion factor and $r_{41}$ is the CO(4--3)-to-CO(1--0) luminosity factor, which depends on the molecular gas excitation (see \citealt{Bolatto13, Carilli13, Pensabene24} for more details). Also in this case we use a lognormal distribution for the prior centred on the CO(4--3) line luminosity $L'_{\mathrm{CO(4-3)}}= (4.2 \pm 0.3) \times 10^{12}\ \mathrm{K \ km/s \ pc^2}$. 

\subsubsection{Stellar bulge}
\label{sec:bulge}
The JWST images at rest-frame wavelengths of 0.4 $\rm \mu m$ to 0.8 $\rm \mu m$ and the light profile presented in W25, show the presence of light excess at the centre of the Big Wheel. Both Chandra X-ray observations and nebular emission-line ratios indicate the presence of an AGN (\citealt{Travascio25, Travascio25b, WangX25}, which could provide a physical explanation for the observed central emission. Nevertheless, we opted for constructing a flexible model and including a bulge since it cannot be excluded by these observations. 

In particular, we assume that the bulge mass distribution follows a Sérsic profile
(see e.g., \citealt{Prugniel97, Terzic05}) of the form:
\begin{equation}
\begin{gathered}
    \rho(r) = \rho_0 \left( \frac{r}{R_{\rm e}}\right)^{-p}
              \exp\left[-b\left(\frac{r}{R_{\rm e}}\right)^{1/n}\right] \text{,}\\
    \rho_0 = \frac{M}{L}I_0 \ b^{n(1-p)}
              \frac{\Gamma(2n)}{2R_{\rm e}\Gamma(n(3-p))} \text{,}
\end{gathered}
\end{equation}
where $r$ is the spatial radius and $\rho_0$ is a normalization factor. The parameter $n$ is the Sérsic index and describes the curvature of the profile. We adopt $p = 1.0 - 0.6097/n + 0.05563/n^2$ from \cite{LimaNeto99}, which provides an accurate approximation to the exact deprojected Sérsic profiles obtained numerically. The term $b$ is a function of $n$ chosen to ensure $R_{\rm e}$ contains half of the galaxy light. A good approximation of $b$ in the range $0.5 <n < 10$ is $2n-1/3+0.009876/n$ (see \citealt{Prugniel97} for more details). In the end, by assuming spherical symmetry, we can derive circular velocity as $v_{\mathrm{circ}}(r)=\sqrt{(GM(r)/r}$, where the enclosed mass can be obtained by following the procedure described in the Appendix A of \cite{Terzic05}:
\begin{equation}
    M(r) = 4 \pi \rho_0 {R_{\rm e}}^3 nb^{n(p-3)}\gamma(n(3-p), \ Z) \text{,}
\end{equation}
where $Z = b (r/R_{\rm e})^{1/n}$ and $\gamma(a,x)$ is the incomplete gamma function. In our model, based on the fit to the observed JWST imaging data at $3 \mu \rm m$, we adopt a fixed value of $n=1.2$, which is more consistent with a pseudo-bulge instead of a classical one ($n=4$). The exact value of $n$ is however not fundamental for determining the DM halo mass. 
Regarding the bulge half-mass radius, we adopt a uniform prior range of $0.1-1$ kpc, consistent with expectations for compact and secularly formed bulges \citep{Shen03, Kormendy04, Gadotti09}. The presence of an AGN, which hampers an accurate determination of the stellar bulge component, prevents tighter constraints on this parameter. Despite the very high spatial resolution of JWST images, performing a reliable morphological analysis of the Big Wheel remains challenging due to its complex structure.

\subsection{Results of the dynamical model}
\label{sec:results_dynmod}
\begin{figure*}[ht!]
    \centering
    \includegraphics[width=\textwidth]{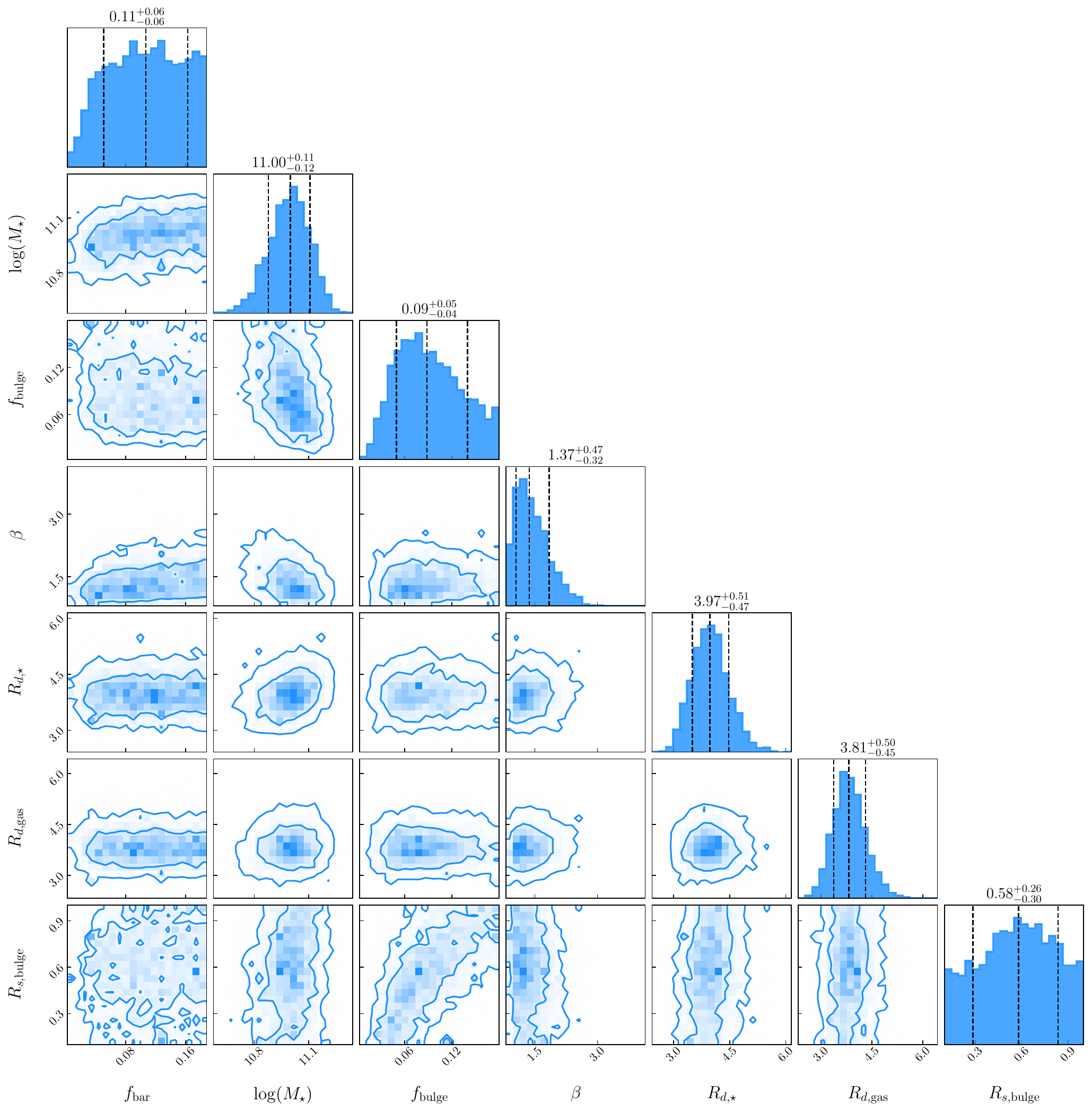} 
    \caption{Best-fit parameters derived from the dynamical model of the Big Wheel galaxy obtained with \texttt{dynesty}, including, from left to right: the baryon fraction $f_{\rm bar}$, the the stellar mass $\log (M_{\star})$, the bulge fraction $f_{\rm bulge}$, the ratio between $\alpha_{\rm CO}$ and $r_{41}$, expressed as $\beta$, the scale radii of both the stellar and gaseous disc, $R_{d, \star}$ and $R_{d, \rm gas}$ respectively, and the scale radius associated with the bulge $R_{s, \rm bulge}$. Given the observational limitations discussed in Section~\ref{sec:overview_data}, the values and the distributions obtained for the scale radii of the gaseous disc and the stellar bulge are mostly driven by the choice of the priors.}
    \label{fig:corner_plot}
\end{figure*}

\newcolumntype{L}{>{\raggedright\arraybackslash}X}
\newcolumntype{Y}{>{\centering\arraybackslash}X}
\begin{table}[!t]
\def\arraystretch{1.3}
\begin{threeparttable}[b]
   \caption{Summary of prior distributions and ranges for the dynamical model parameters of the Big Wheel galaxy.}
\label{tab:priors}
\centering
\begin{tabularx}{0.99\columnwidth}{L Y Y}
\toprule
\toprule
Parameter & Prior distribution & Range \\
\midrule
\midrule
$\log(f_{\mathrm{bar}})$ & uniform & $[-4\text{,} \ \log(f_{\rm bar, \rm c})]$ \tnote{a} \\
$\log (\rm M_{\star}/\msun)$ & gaussian & $11.37^{+0.20}_{-0.20}$ \\
$\log(f_{\rm bulge})$ & uniform & $[10^{-3}, \ 0.18]$ \\
$\alpha_{\rm CO} \rm(\rm\msun/K \ km \ s^{-1} \ pc^{-2})$ & uniform & [0.8, 4.3] \\
$r_{41}$ & uniform & [0.2, 1] \\
$R_{d, \star}$ (kpc) \tnote{(b)} & lognormal & $3.75^{+0.48}_{-0.36}$ \\
$R_{d, \rm gas}$ (kpc) \tnote{(b)} & lognormal & $3.75^{+0.48}_{-0.36}$ \\
$R_{s, \rm bulge}$ (kpc) & uniform & [0.1, 1]  \\
\bottomrule
\bottomrule
\end{tabularx}
\begin{tablenotes}
       \item [(a)] $f_{\rm bar, \rm c} \approx 0.187$.
       \item [(b)] $R_d=r_{\rm half\text{-}mass}/1.678$, where $r_{\rm half\text{-}mass} = 6.3^{+0.8}_{-0.6}$ kpc.
     \end{tablenotes}
  \end{threeparttable}
\end{table}

\newcolumntype{L}{>{\raggedright\arraybackslash}X}
\newcolumntype{Y}{>{\centering\arraybackslash}X}
\begin{table}[!t]
\def\arraystretch{1.3}
\begin{threeparttable}[b]
   \caption{Dynamical model parameters (upper table) and derived quantities (lower table) of the Big Wheel galaxy. The results are reported as median values and corresponding $68\%$ uncertainties.}
\label{tab:posteriors}
\centering
\begin{tabularx}{0.99\columnwidth}{L Y Y}
\toprule
\toprule
Best-fitting parameters & Posteriors \\
\midrule
\midrule
$f_{\mathrm{bar}}$ & $0.11^{+0.06}_{-0.06}$\\
$\log (\rm M_{\star}/\msun)$ & $11.00^{+0.11}_{-0.12}$  \\
$f_{\rm bulge}$ & $0.09^{+0.05}_{-0.04}$ \\
$\beta$ ($\rm \msun/K \ km \ s^{-1} \ pc^{-2}$) \tnote{a} & $1.37^{+0.47}_{-0.32}$\\
$R_{d, \star}$ (kpc) & $3.96^{+0.51}_{-0.47}$ \\
$R_{d, \rm gas}$ (kpc) & $3.81^{+0.50}_{-0.44}$ \\
$R_{s, \rm bulge}$ (kpc) & $0.58^{+0.26}_{-0.29}$  \\
\midrule
\midrule
\\
\midrule
\midrule
Derived parameters & Values \\
\midrule
\midrule
$\log (\rm M_{DM}/\msun)$ & $12.11^{+0.29}_{-0.17}$  \\
$c_{\rm DM}$ & $2.97^{+0.80}_{-0.63}$  \\
$\log (\rm M_{\star,disc}/\msun)$ & $10.95^{+0.12}_{-0.13}$  \\
$\log (\rm M_{\star, bulge}/\msun)$ & $9.93^{+0.19}_{-0.23}$  \\
$\log (\rm M_{gas}/\msun)$ & $10.76^{+0.13}_{-0.12}$  \\
\bottomrule
\bottomrule
\end{tabularx}
\begin{tablenotes}
       \item [(a)] $\beta=\alpha_{\rm CO}/r_{41}$.
     \end{tablenotes}
  \end{threeparttable}
\end{table}

\begin{figure}
    \centering
    \includegraphics[width=\columnwidth]{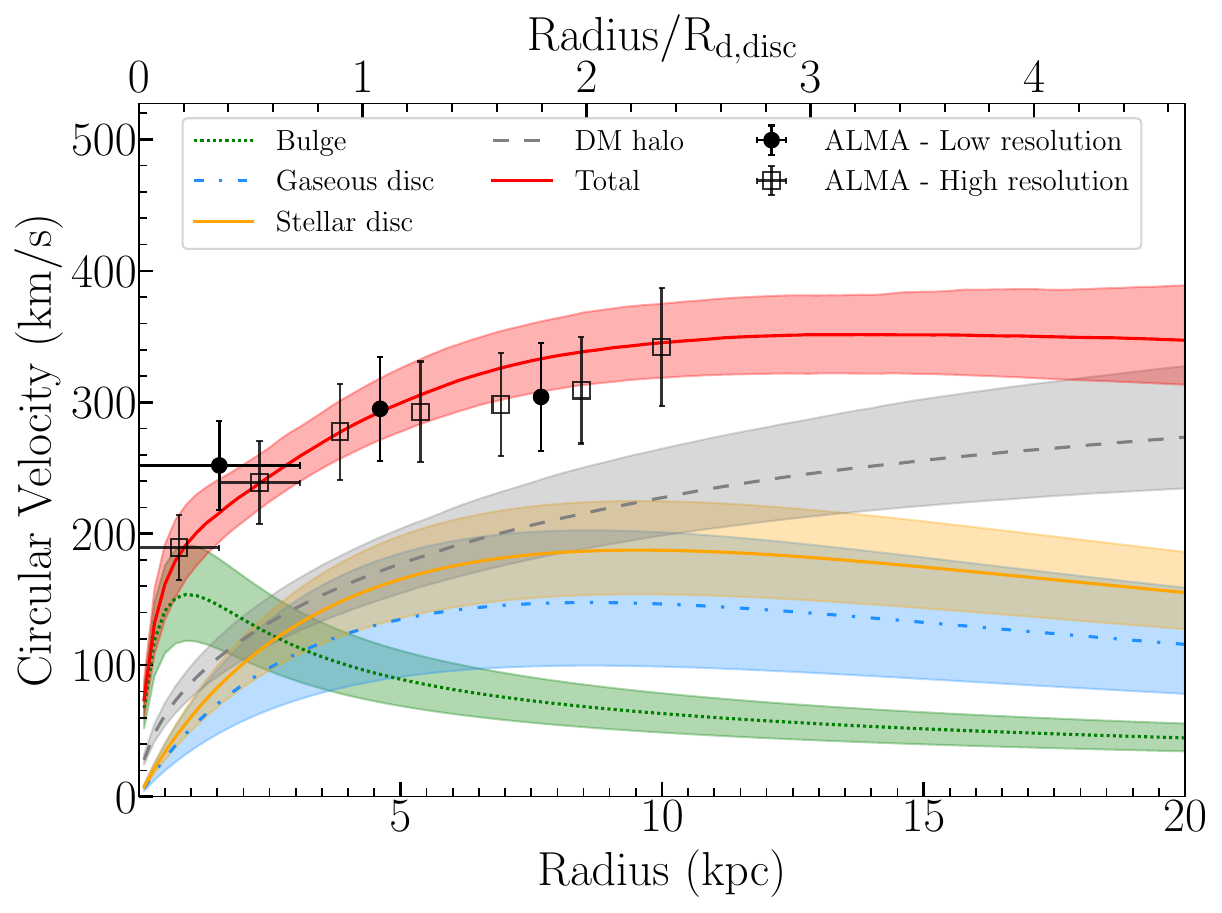} 
    \caption{Velocity curve of the Big Wheel derived from the best-fit parameters presented in Figure~\ref{fig:corner_plot} showing the different contributions of the galaxy separate components (see legend) as a function of galactocentric radius.
    The shaded areas indicate values between the $16^{\rm th}$ and $84^{\rm th}$ percentiles. The circles and open squares represent the ALMA CO(4--3) low-resolution and high-resolution data, respectively, used in the analysis (see Section~\ref{sec:highres_data}).
    The bars on the x-axis for the first point of the low-resolution dataset and for the first two points of the high-resolution data are shown in order to illustrate the different spatial binning of the two datasets.}
    \label{fig:vrotCO}
\end{figure}

The results of the Bayesian analysis conducted with \texttt{dynesty} are reported in Figure~\ref{fig:corner_plot} as a corner plot and summarised in Table~\ref{tab:posteriors}. In particular, we find a DM halo mass of $\log(\rm M_{DM}/\msun)=12.11^{+0.29}_{-0.17}$, with a concentration $c_{\rm DM}=2.97^{+0.80}_{-0.63}$. The total stellar mass is $\log(\rm M_{\star}/\msun)=11.00^{+0.11}_{-0.12}$, distributed between a stellar disc with $\log(\rm M_{\star, disc}/\msun)=10.95^{+0.12}_{-0.13}$ and a bulge with $\log(\rm M_{\star, bulge}/\msun)=9.93^{+0.19}_{-0.23}$. Respectively, the associated scale radii are $R_{d, \star}=3.96^{+0.51}_{-0.47}$ kpc and $R_{s, \rm bulge}=0.58^{+0.26}_{-0.29}$. Finally, the gas mass is estimated to be $\log(\rm M_{gas}/\msun)=10.76^{+0.13}_{-0.12}$, with a scale radius of $R_{d,\rm gas}=3.81^{+0.50}_{-0.44}$ kpc. In Figure~\ref{fig:vrotCO} we show the velocity curve decomposition for the Big Wheel, derived from the median of the best-fit parameter distributions. The different lines correspond to the profiles of the different components, described in Section~\ref{sec:dyn_mod}, where the dashed grey line corresponds to the velocity curve of the DM halo, the solid orange line to the one related to the stellar disc, the dash-dotted blue line to the gaseous disc one, and the dotted green line to the bulge one. The solid red line represents the velocity curve of the total system. For all curves, the shaded areas represent the values between the $16^{\rm th}$ and $84^{\rm th}$ percentiles.
The black circles represent the rotation curve velocities extracted from the low resolution rotation map of the ALMA CO(4--3). The open squares show the data extracted from the high-resolution ALMA CO(4--3) rotation map as discussed in Section~\ref{sec:highres_data}. The error bars contain the uncertainties on the galaxy inclination. 

An additional relevant aspect to consider is that the plot in Figure~\ref{fig:vrotCO} is constructed from the posterior distribution of the model parameters. The solid curve represents the $50^{\rm th}$ percentile (median) of the posterior, computed using the normalized importance weights from nested sampling, where each sample is weighted proportionally to its likelihood times the associated prior volume element, $w_i\propto \mathcal{L}_i\Delta X_i$. The shaded regions indicate the $16^{\rm th}$ and the $84^{\rm th}$ percentiles, and thus encompass the full posterior volume, including regions that do not correspond to the maximum-likelihood solution. To evaluate the impact of this effect, we conducted an additional analysis restricted to the $20$ parameter sets exhibiting the highest likelihood values, as determined by the Bayesian sampling procedure. The median values for the DM halo and the stellar masses, computed from the parameter sets of the subsample, are $\log(\mathrm{M}_h/\msun) = 12.14$ and $\log(\mathrm{M}_\star/\msun) = 11$, respectively. These values are consistent with the case considering the full posterior volume. 

\begin{figure}
    \centering
    \includegraphics[width=\columnwidth]{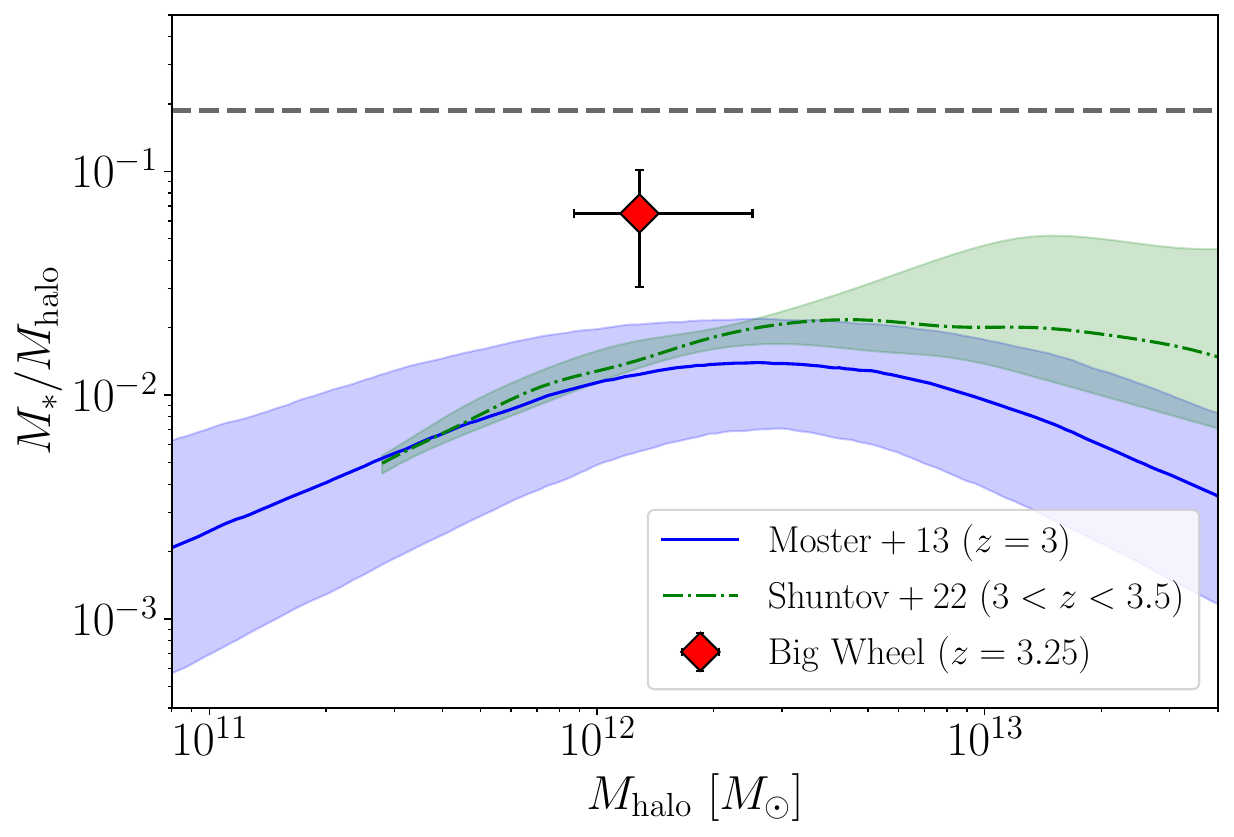} 
    \caption{Derived stellar-to-halo mass (SHM) ratio for the Big Wheel (red diamond with $1\sigma$ uncertainties along both axes), compared with the expected stellar-to-halo-mass relation (SHMR) from \citet{Moster13} at $z=3$ (solid blue line) and \citet{Shuntov22} (dash-dotted green line; including only central halos). The errorbars on the Big Wheel point are obtained from the analysis presented in Figure~\ref{fig:corner_plot}, while, the blue and green shaded areas represent the $16^{\rm th}$ to $84^{\rm th}$ percentiles obtained as discussed in Sec ~\ref{sec:results_dynmod}. This figure shows that the SHM ratio of the Big Wheel is more than $1\sigma$ away from the expected relations. The implication of this result are discussed in Section~\ref{physical_implications}.}
    \label{fig:SHMR}
\end{figure}

In Figure~\ref{fig:SHMR}, we show the resulting SHM ratio for the Big Wheel. The dashed dark grey line corresponds $f_{\rm bar, c} = \Omega_{b}/(\Omega_{m} -\Omega_{b})$. The red diamond represents the measure for the Big Wheel, compared with the multi-epoch abundance matching model by \cite{Moster13}, represented as a solid blue line following the expression:
\begin{equation}
    \frac{M_{\star}}{M_{h}} = 2N \left[ \left(\frac{M}{M_1}\right)^{-\beta} + \left(\frac{M}{M_1}\right)^{\gamma}\right]^{-1}
\end{equation}
where $M_1$, $\beta$, $\gamma$, and the normalisation factor $N$ are all functions of the redshift $z$. The shaded area shows the $68\%$ range of model predictions, obtained from Monte Carlo sampling of the parameter uncertainties. Together with this model, we also plot the expected SHMR relation for central halos at $3 < z < 3.5$ from \citet{Shuntov22} as a dash-dotted green line. The main difference with \cite{Shuntov22} is that they considered a single dataset to probe the SHMR, while \cite{Moster13} relies on observables from heterogeneous data sets, with different functions and methods to select galaxies' properties. In particular, we note that the derived halo mass for the Big Wheel is at the peak of the theoretical model by \cite{Moster13} and that its corresponding SHM ratio deviates by approximately  $1\sigma$ from both the solid blue line representing the relation by \cite{Moster13} and from the dash-dotted green line representing the \cite{Shuntov22} relation. In Section~\ref{physical_implications}, we discuss the implication of these results for our understanding of the formation history of the Big Wheel galaxy.

\section{Numerical Simulations}
\label{sec:num_sim}
\begin{figure*}[h]
    \centering
    \includegraphics[width=\textwidth]{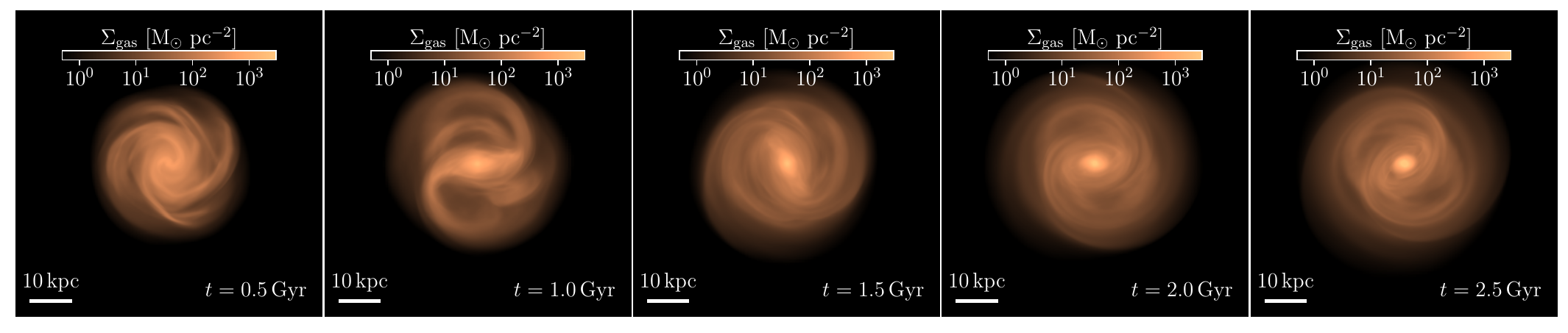}
    \includegraphics[width=\textwidth]{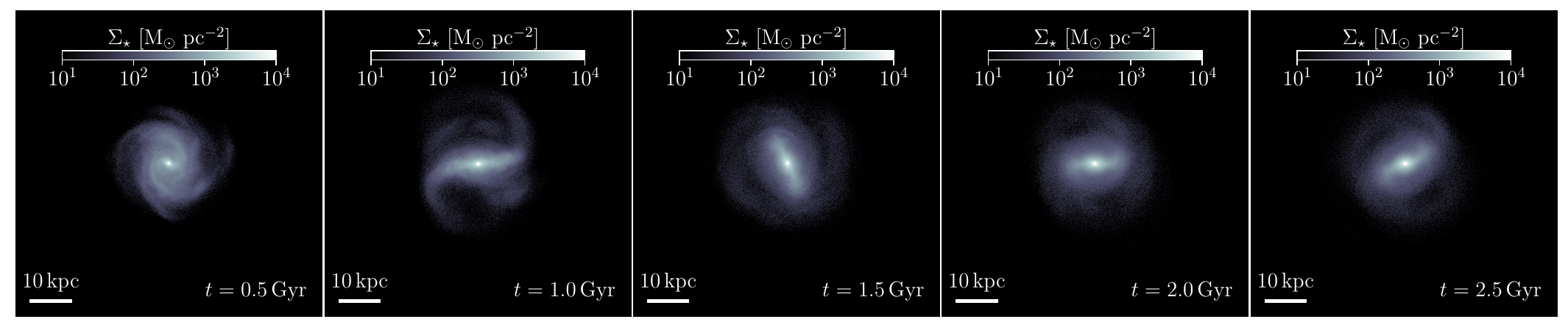}
    \caption{Numerical simulation of an idealized galaxy built from the best-fit parameters derived for the Big Wheel and presented in Figure~\ref{fig:corner_plot}. The top and the bottom panels, show, respectively, the gaseous and the stellar surface density maps of the galaxy, shown face-on, at five different times during the simulation ($t = 0.5$, $1$, $1.5$, $2$, and $2.5$ Gyr). Throughout the entire time evolution, both the gas and the stellar distributions do not exhibit the onset of any global gravitational instabilities. In particular, spiral arms are continuously formed and disrupted, giving rise to substructures like a bar and a pseudo-bulge during the evolution. However, the global disc structure remains preserved.}
    \label{fig:maps_faceon}
\end{figure*}
\begin{figure*}[h]
    \centering
    \includegraphics[width=\textwidth]{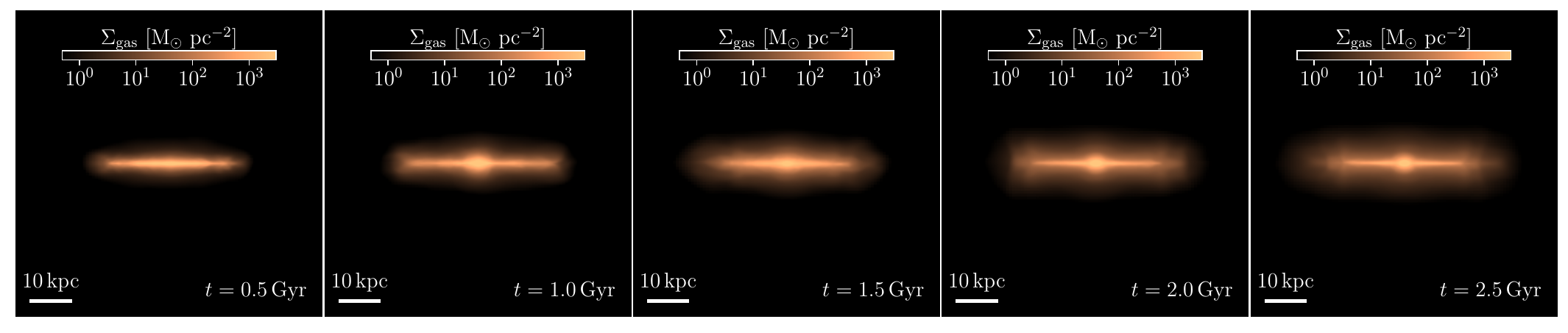}
    \includegraphics[width=\textwidth]{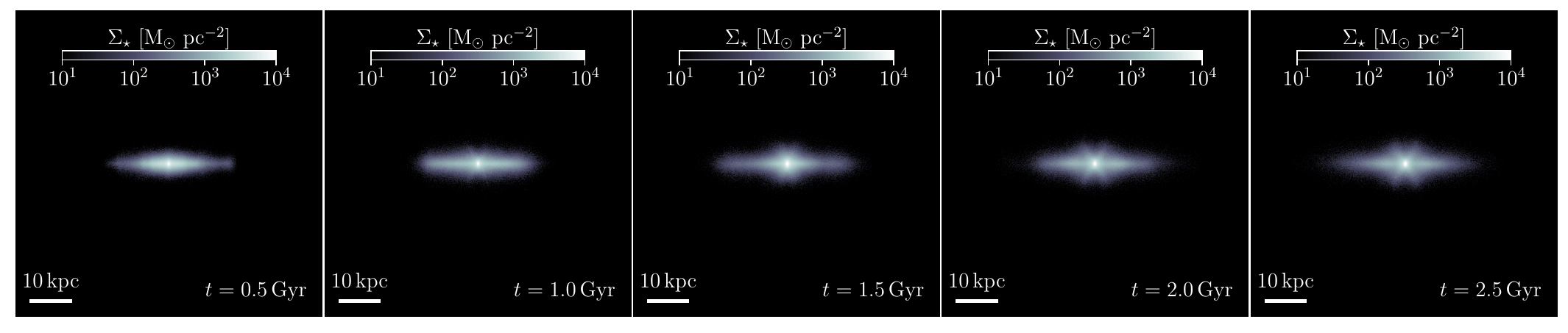}
    \caption{As in Figure~\ref{fig:maps_faceon}, showing instead the edge-on view of the simulated galaxy.}
    \label{fig:maps_sideon}
\end{figure*}

\begin{figure}[h]
    \centering
        \includegraphics[width=\linewidth]{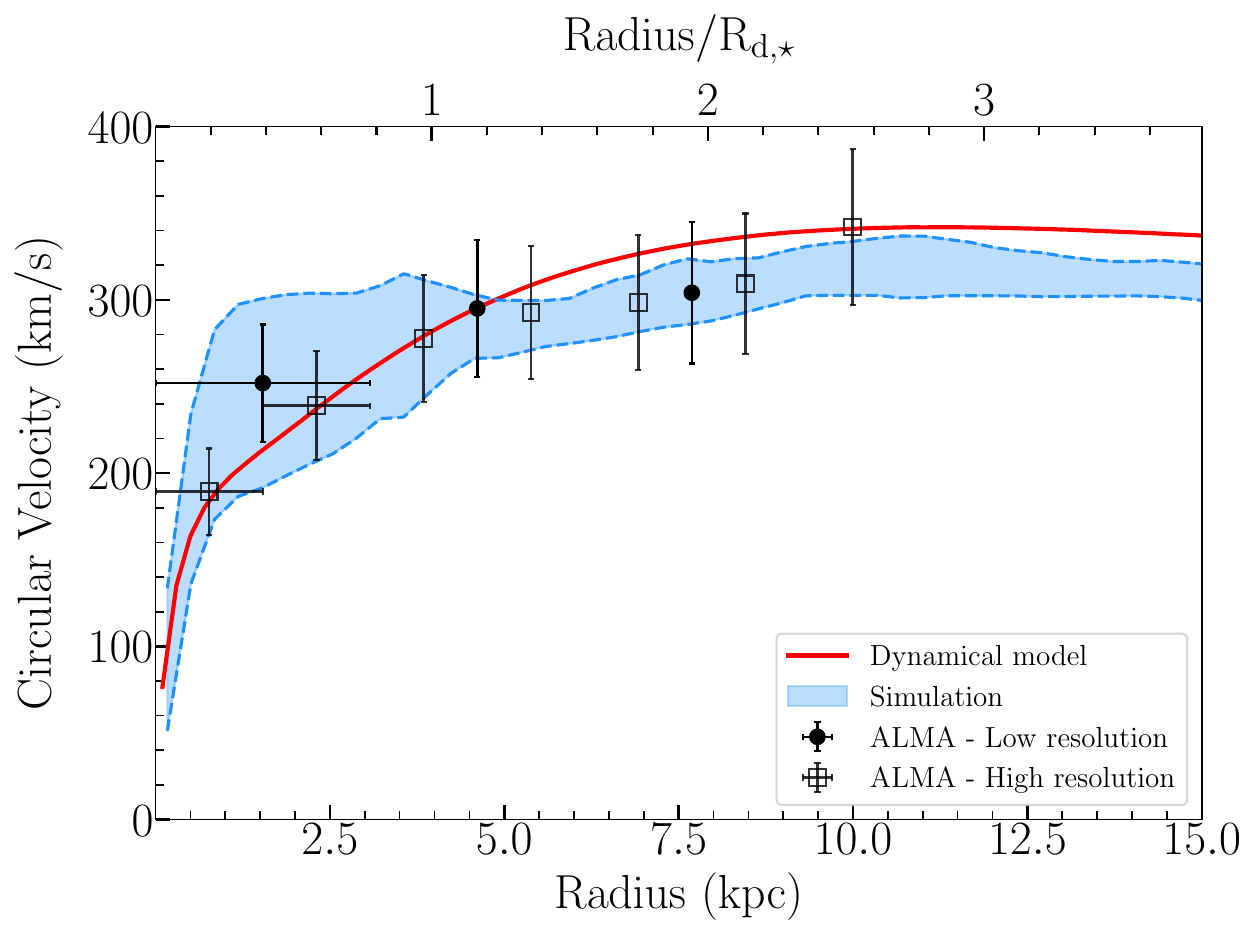}
\caption{Time evolution of total circular velocity of the simulated galaxy. The shaded blue region indicates the range of circular velocities measured between $0.5$ and $2.5$ Gyr. At each radial bin, the boundaries correspond to the minimum and the maximum values across the time interval. The envelope is compared to the initial conditions (red solid line) and the observational datasets (see Section \ref{sec:almaC0}).}
\label{fig:circvel_sim}
\end{figure}

We complement the analysis above by testing that the resulting galaxy is unaffected by gravitational instabilities, which could make it substantially different from the Big Wheel galaxy as currently observed. To assess whether such instabilities could develop, we performed numerical simulations of an idealised system using the adaptive mesh refinement (AMR) hydrodynamical code \ramses{} \citep{Teyssier02} in order to verify the general stability of the galaxy for at least $3$ dynamical times. 

The galaxy initial conditions (ICs) are generated through the software DICE (Disc Initial Conditions Environment) by \cite{Perret14} using the best-fit parameters obtained by \dynesty. The galaxy is then evolved in isolation and adiabatically. To ensure stability, we set a minimum initial Toomre parameter for the stellar disc of $Q_{\star,\mathrm{min}}=3$. The refinement level of the coarse gravitational potential grid is 7, and it is refined up to level 12. At the highest refinement level, the spatial resolution of the potential grid is equal to $x_0 = 100$ pc, corresponding to the minimum resolved spatial scale of the gravitational potential. The mass resolution of each component is equal to $10^6 \ \msun$ for the DM halo and $10^5  \ \msun$ for all baryonic components, i.e. stellar disc, the bulge, and the gas. We did not impose any a priori equilibrium, neither hydrostatic nor thermal. For each component, we imposed the models described in ~\ref{sec:dyn_mod} for the density distribution. In addition, we assume a gas temperature of $4\times10^4$ K.

Simulations are performed within a computational domain with size $l_{\rm box} = 400$ kpc. The coarsest level of the AMR grid is $\ell_{\rm min} = 7$, corresponding to a cartesian grid with $(2^7)^3$ cells with a size of $\Delta x=3.125$ kpc. The finest resolution level of the grid is $\ell_{\rm max}=12$, corresponding to a cell size of $\Delta x = 100$ pc. The system is then evolved for a total time duration of $\Delta t =2.5$ Gyr
which corresponds approximately to $15$ dynamical timescales $t_{\rm dyn}=2\pi R/V_\mathrm{c}\approx0.16$ Gyr. 

In Figure~\ref{fig:maps_faceon} we show five snapshots at different times ($t = 0.5$, $1$, $1.5$, $2$, and $2.5$ Gyr) representing the gas (top panel) and stellar (bottom panel) surface density as seen face-on. Neither distribution exhibits evidence of global instabilities that could lead to a significant expansion or compression of the galaxy. 
Features resembling spiral arms are present in both the stellar and gas spatial distributions. The innermost region shows an elongated shape, consistent with the presence of a bar evolving towards a pseudo-bulge. This structure is especially visible in Figure~\ref{fig:maps_sideon}, which shows the surface density maps for both gas and stars as seen edge-on.

In Figure~\ref{fig:circvel_sim}, we present the circular velocity curve of the simulated galaxy over the time interval between $0.5$ and $2.5$ Gyr. With respect to the corresponding dynamical model (solid red line), a localised excess appears at radii up to $5$ kpc, clearly associated with the development of a bar evolving into a pseudo-bulge that have significantly altered the mass distribution, as shown in Figures~\ref{fig:maps_faceon} and~\ref{fig:maps_sideon}. Furthermore, orbits within bars are typically non-circular. As a result of the significant concentration of matter within the innermost region, the gravitational potential becomes deeper in the centre, leading to a relaxation of the curve at larger radii, dropping below the initial dynamical model. In this regard, we have further verified that, at large radii, the simulation is in good agreement with the H$\rm \alpha$ line-of-sight velocity ($v_{\rm los}$) profile extracted from the JWST/NIRSpec slit closer to the centre of the Big Wheel galaxy (W25). Additional details will be presented in Pensabene et al., in prep.).

\subsection{Disc Stability}
The results of our numerical experiments in adiabatic conditions, presented in Figure~\ref{fig:maps_faceon} and Figure~\ref{fig:maps_sideon}, show that a galaxy characterized by the best-fit parameters of our dynamical model does not develop global instabilities – neither contraction nor significant expansion. Moreover, the radii of the stellar and gas distributions remain approximately unchanged during the simulation time evolution \citep{Fangzhou25}. 
We stress that the disc gas temperature has been initialized in such a way to ensure initial stability. However, this choice does not have an impact on the stability of the stellar disc, since its mass is four times higher, and it is thus the major dynamical component of the system. 

Interestingly enough, the stellar surface density exhibits signatures of spiral arms throughout the entire evolutionary sequence, while a roughly spherical structure associated with the central bulge persists across the different snapshots. Toward the end of the simulation, this central feature becomes slightly elongated, with a phase in which the spiral arms appear to open. These morphological features could be used in future studies to provide additional physical constraints by a more quantitative comparison with the observational data. In particular, they could be used to better constrain the matter distribution in terms of non-axisymmetric features within the disc and within the innermost region, including the effect of a MBH \citep{Bonoli16}.

\section{Discussion}
\label{sec:discussion}
\subsection{Physical implications}
\label{physical_implications}
The relation between the stellar and DM halo masses provides a fundamental insight into the efficiency at which a galaxy has formed stars and its more recent evolutionary history. In this context, an important result of our analysis is that the Big Wheel appears to have been very efficient in forming stars, as it is shown by its SHM ratio value of $M_\star/M_h=0.06^{+0.05}_{-0.03}$. This value is approximately three times larger than the expectation from abundance-matching studies, as presented in Figure~\ref{fig:SHMR}. 
As ejective feedback is usually considered one of the key processes influencing the SHM ratio of a galaxy of that mass (\citealt{FaucherGiguere11,Behroozi13,Somerville15}), our result suggests that the Big Wheel has likely not experienced strong stellar or AGN ejective feedback episodes that permanently removed the majority of the gas during its formation history \citep{Posti21}. Episodic feedback may still have occurred, with expelled gas either remaining bound or later reaccreted via galactic fountains. Indeed, the stellar mass of the Big Wheel is about half the expected maximum baryonic content of the entire halo associated with this galaxy which should include also the galaxy's ISM and its CGM. Although we do not have detailed constraints on the total mass associated with these components, the SFR of the Big Wheel, which is consistent with main-sequence galaxies, suggests no lack of gas supply in its ISM and CGM. As such, it is not implausible to imagine that the overall baryonic content of the Big Wheel halo could be consistent with the one expected from the cosmological baryonic fraction. 

Such a high SHM ratio of the Big Wheel as estimated through our dynamical model is at odds with the expectation from the general galaxy population at $z\sim3$ produced by halo occupation distribution (HOD)-based models using different datasets and assumptions, such as the one by \cite{Moster13} and \cite{Shuntov22} as shown in Figure~\ref{fig:SHMR}. We note that the SHMR profile by \cite{Shuntov22} is extrapolated for DM halo masses above $2\times 10^{12} \msun$. Instead, the Big Wheel SHM ratio is more similar to the SHMR of local super-spiral galaxies, a very rare population of local massive disc galaxies in which star formation has not been yet suppressed \citep{Ogle16, Ogle19, DiTeodoro21, DiTeodoro23}. 
This is consistent with other peculiar properties of the Big Wheel, such as its stellar half-light radius (and half-mass radius), which has been found to be similar to the one of local super-spiral galaxies at the same stellar mass instead of following the expected galaxy radius-mass relation at $z\sim3$ (see W25 for details). 
The location of the Big Wheel within one of the largest overdensities of galaxies known at this redshift (\citealt{Pensabene24, Pensabene25, Galbiati25, Travascio25, Travascio25b}) could suggest a relation between these peculiar properties and the environment.

\begin{figure}
    \centering
    \includegraphics[width=\columnwidth]{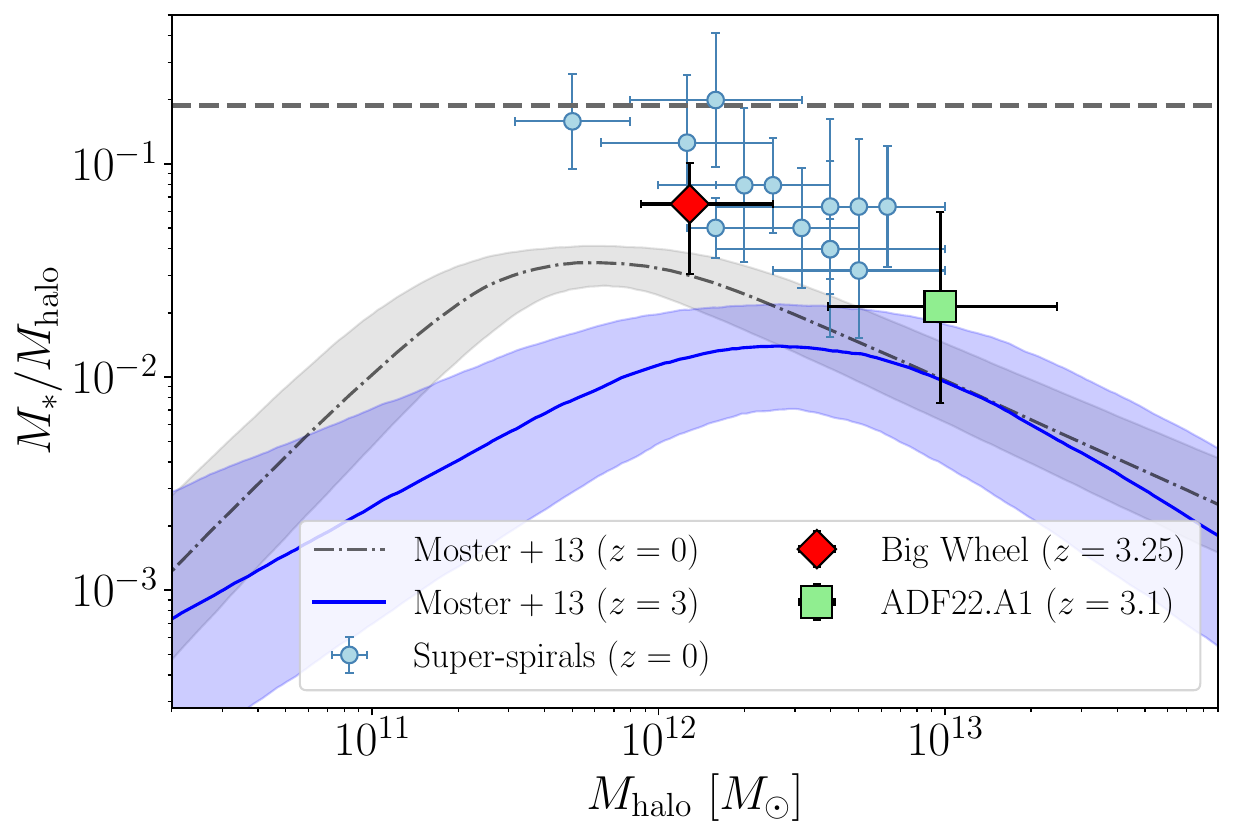} 
    \caption{Comparison between the SHM ratio of the Big Wheel (red diamond) with the other known large disk galaxy at similar redshift (ADF22.A1; green square) as obtained by our model (see Appendix~\ref{appendix:adf22.a1}) and with super-spiral galaxies in the local universe (blue circles), as compiled by \cite{DiTeodoro23}. We note that, within the local sample of galaxies, we have excluded two systems (NGC5635 and UGC12591) for which robust DM halo constraints are not available, following the approach adopted in \cite{DiTeodoro23}. The expected SHMR at $z=3$ and in the local universe from \cite{Moster13} are shown as solid and dot-dashed lines, respectively. The uncertainties in the expected relations is obtained as in Figure~\ref{fig:SHMR}. 
    As for its relation between mass and size (see W25), the Big Wheel presents a SHM ratio similar to the local super-spiral galaxies. The implication of this result are discussed in Section~\ref{physical_implications}. 
    }
    \label{fig:SHMR_bw_adf22.a1}
\end{figure}

To date, only one other massive disc galaxy within exceptional overdensities ($\delta\gg1$) has been reported with spectroscopic confirmation and kinematic measurements in the literature, i.e. ADF22.A1 within the SSA22 proto-cluster \citep{Umehata25}. This galaxy has a stellar mass ($M_{\star}=2.51\times10^{11} \ \msun$) and half-light radius ($7$ kpc) comparable to those of the Big Wheel. However, ADF22.A1 has a much higher rotation velocity than the Big Wheel ($v_{\rm rot} = 331 \pm 30 \ \mathrm{km \ s^{-1}}$), reaching a value of $v_{\rm rot} = 530 \pm 10 \ \mathrm{km \ s^{-1}}$. We applied the same procedure used for the Big Wheel and described in Sect ~\ref{sec:dyn_mod} to constrain both the DM mass and concentration of the ADF22.A1 galaxy halo. The complete set of results is presented in Appendix~\ref{appendix:adf22.a1}, with all the assumptions made for the priors and the results for the posterior distribution. Figure~\ref{fig:SHMR_bw_adf22.a1} shows the SHM ratio from both the Big Wheel (red diamond) and the ADF22.A1 galaxy (light green square), both with $1\sigma$ uncertainties along both axes. It is evident that, whereas the Big Wheel lies approximately $2\sigma$ from the SHMR of \cite{Moster13} at $z=3$ (solid blue line), the data point associated with the ADF22.A1 galaxy corresponds to substantially bigger DM halo, with a mass of the order of $10^{13} \ \msun$. As a result, the system has a SHM ratio which is consistent with the SHMR of \cite{Moster13} within the $1\sigma$ uncertainty, and located at the edge of the region occupied by the local super-spiral sample. As such, ADF22.A1 could represent a system with a different origin with respect to the Big Wheel. Unfortunately, there is currently no statistical sample of massive disc galaxies located within large overdensities at $z\sim3$. In this regard, a similar result has been recently presented in \cite{Rizzo26}, although we note that their dynamical analysis relies on a different kinematical model than the one adopted here, which is based on the original data provided by \cite{Umehata25}.

\subsection{Spin parameter and specific angular momentum}
\label{sec:spin_jk}
In this context, one possible explanation for the exceptionally large size of this galaxy is that the spin parameter of its host halo is sufficiently high to enhance the angular momentum of the galaxy. 
According to Eq. (28) in \cite{Mo98}, the spin parameter $\lambda$ can be expressed as:
\begin{equation}
\label{eq:lambda}
    \lambda=\sqrt{2}\frac{R_d}{r_{200}}\frac{m_d}{j_d}f_c^{1/2}f_R(\lambda, c, m_d, j_d)^{-1}
\end{equation}
where $m_d$ and $j_d$ denote, respectively, the mass and angular momentum of the baryonic disc component, expressed as fractions of the DM halo mass and angular momentum (i.e. $M_{\rm bar}=m_dM_{\rm DM}$ and $J_{\rm bar}=j_dJ_{\rm DM}$). $R_d$ is the scale radius of the disc, $r_{200}$ is the virial radius of the DM halo, and $f_c$ is a function of the concentration parameter $c$, which in our case can be approximated to unity. The factor $f_R$ is given by:
\begin{equation}
    f_R(\lambda, c, m_d, j_d)=2\left[ \int_{0}^{\infty} e^{-u}u^2\frac{V_c(R_du)}{V_{200}}\,du\right]^{-1}
\end{equation}
where $V_c(R)$ is the circular velocity at radius $R$ and $V_{200}$ is the circular velocity at the virial radius $r_{200}$. \

From our dynamical model, we have the functional form of $V_c(r)$, so we can directly calculate $f_R$, which also in this case is close to unity. We approximate $m_d$ with the value of the SHM ratio of the Big Wheel, i.e. $m_d=0.06$. However, we have no information on the specific angular momentum of the dark matter component and thus on $j_d$. The standard assumption adopted in models of galaxy formation (e.g., \citealt{Mo98, Somerville99, Cole00}) is to consider the ratio
between the specific angular momentum of disc galaxies and their host dark matter haloes equal to $\mathcal{R}_j=j_d/m_d=1$. In this regard, the super-spiral galaxies studied by \cite{DiTeodoro21} appear to have conserved their angular momentum, exhibiting a ratio between the specific angular momentum of the baryons and the DM halo consistent with unity. Making this assumption would result into $\lambda=0.085$. Other works suggest however a value of $\mathcal{R}_j<1$, i.e. some angular momentum loss during galaxy formation. For instance, \cite{Dutton12} combined constraints on the galaxy-DM connection with structural and dynamical scaling relations to investigate the angular momentum content of disc galaxies, finding $\mathcal{R}_j=0.61^{+0.13}_{-0.11}$ which would result into $\lambda=0.139$. In the reminder of this Section, we will consider these two values as possible limits for the Big Wheel halo:
\begin{equation}
    \lambda_\text{BW}= \left\{ \begin{array}{rcl}
{0.085} & \mbox{for} &  \mathcal{R}_j= 1 \\ 
{0.139} & \mbox{for} & \mathcal{R}_j=0.61
\end{array}.\right.
\end{equation}
\cite{Maccio} investigated the structural properties of dark matter haloes using a statistical sample of halos drawn from a series of N-body simulations, finding that the distribution of the halo spin parameters can be fitted with a lognormal distribution that peaks around $\lambda_0=0.034\pm0.001$ with $\sigma_{\log{\lambda}}=0.55\pm0.01$, meaning that our estimates for the spin parameter of the Big Wheel are around $2\text{-}4$ times the typical expected value considering $\mathcal{R}_j= 1$ or $\mathcal{R}_j= 0.61$ respectively. This result is in line with the size measurements of the Big Wheel, which show a scale radius of the stellar disc three times larger than expected for typical galaxies of similar mass and redshifts (see W25, for further discussion). We performed the same analysis for the ADF22.A1 galaxy, obtaining the following results for the spin parameter of its DM halo:
\begin{equation}
    \lambda_{\text{ADF22.A1}}= \left\{ \begin{array}{rcl}
{0.038} & \mbox{for} &  \mathcal{R}_j= 1 \\ 
{0.063} & \mbox{for} & \mathcal{R}_j=0.61
\end{array} . \right. 
\end{equation}
The result suggests that ADF22.A1 resides in a massive DM halo with a lower spin parameter with respect to the Big Wheel one. This implies that the large spatial extent of the galaxy could be primarily driven by the mass and scale of its host halo rather than by its high rotational support. Our dynamical analysis thus confirms that a viable mechanisms to produce a galaxy as large as the Big Wheel could require a halo with a value of $\lambda$ which is at least $2\sigma$ away from the expectations for typical haloes \citep{Fangzhou25}. Alternatively, if the $\lambda$ of the Big Wheel halo is similar to the typical value found by \cite{Bullock01}, then our analysis would require a value of $\mathcal{R}_j\sim2$, meaning that the baryonic component of the Big Wheel should have a specific angular momentum twice as large as the one of its parent DM halo.

Our estimates for the spin parameter are in excellent agreement with the methodology described in \cite{DiTeodoro23}, where they perform a complementary analysis: by fixing the spin parameter at the peak value of its distribution ($\lambda \approx0.034$), they recover the specific angular momentum of the galactic components, normalized to that of the DM. This is indicated as $f_j$, but is physically the same quantity as $\mathcal{R}_j$ described above, this time derived from the data under the assumption that $\lambda = 0.034$. In the following, we adopt this approach. The specific angular momentum $j_{k}$ for a given component $k$ (stars or gas) modelled as an exponential disc is:
\begin{equation}
\label{eq:fall_eq}
    j_k(<R) = \frac{J_k(<R)}{M_k(<R)}=\frac{\int_{0}^{R} \Sigma_k(R'){R'}^{2}V_{\rm rot}(R')\,dR'}{\int_{0}^{R} \Sigma_k(R'){R'}\,dR'},
\end{equation}
where $\Sigma_k(R) = M_k / (2 \pi R_d^2) \exp(-R / R_d)$ is the surface mass density with the scale radius $R_d$, and $V_{\rm rot}(R)$ is the rotational velocity, which we assume to be equal to the circular velocity at all radii for all components. To derive the posterior distribution of the specific angular momentum of stars, $j_\star$, and gas, $j_{\rm gas}$, we compute the integral in Eq.(\ref{eq:fall_eq}) for each individual sample of the \dynesty{} chain. This procedure allows us to propagate the uncertainties and degeneracies directly into the angular momentum estimates. The final values and their associated $1\sigma$ uncertainties correspond to the median and the  $16^{\rm th}$-$84^{\rm th}$ percentile range of the derived posterior distribution. The specific angular momentum of the baryons is $j_{\rm{bar}} = f_{\rm{gas}}j_{\rm{gas}} + (1-f_{\rm{gas}})j_{\rm{\star}}$, where $f_{\rm{gas}}=M_{\rm{gas}}/M_{\rm{bar}}=M_{\rm{gas}}/(M_{\rm{gas}}+M_{\rm{\star}})$. 
To account for the contribution of the galaxy outskirts, where a substantial fraction of the angular momentum is expected, we extend the integration up to $R=20$ kpc and $R=35$ kpc instead of up to the observed radius only ($R_{\rm obs}\approx10$ kpc), assuming that the gas and stellar mass distribution follow an exponential profile and that the rotation curve follows the dynamical model (and therefore remains flat, see Figure \ref{fig:vrotCO}) also at large radii. As for the specific angular momentum of the DM halo, according to \cite{Bullock01}, this is equal to
\begin{equation}
\label{j_dm}
    j_{\rm{ DM}}=\lambda\sqrt{2} R_{200}V_{200},
\end{equation}
where $R_{200}$ is the radius of the DM halo containing the virial mass $M_{200}$ and $V_{200}$ is the circular velocity of the DM halo at the corresponding $R_{200}$. The spin parameter $\lambda$ is assumed equal to the value of the peak of the lognormal distribution derived in \cite{Maccio}, i.e. $\lambda=0.034$. The stellar-to-halo specific angular momentum ratio is $f_{j_{\star, \rm BW}}=2.03^{+0.60}_{-0.75}$ and the one for the total baryonic content is $f_{j_{\rm bar, BW}}=2.03^{+0.60}_{-0.74}$. These values are similar to the highest values found in sample of super-spiral galaxies in the local Universe studied by \cite{DiTeodoro23}. We performed the same analysis for the ADF22.A1 galaxy, finding that the specific angular momentum ratio for the stellar disc is $f_{j_{\star, \rm ADF22.A1}}=0.91^{+0.74}_{-0.42}$ and the one for the total baryonic content is $f_{j_{\rm bar, ADF22.A1}}=0.92^{+0.74}_{-0.43}$. The results are in agreement with what found in \cite{Rizzo26} for both quantities. The specific angular momenta of all components of both the Big Wheel and ADF22.A1 are presented in Table~\ref{tab:j_k} and Table~\ref{tab:j_k_adf22.a1} (see Appendix~\ref{appendix:adf22.a1}).

\newcolumntype{L}{>{\raggedright\arraybackslash}X}
\newcolumntype{Y}{>{\centering\arraybackslash}X}
\begin{table}[!t]
\def\arraystretch{1.3}
\begin{threeparttable}[b]
   \caption{Specific angular momentum for all components of the Big Wheel. The specific angular momentum of the DM $j_\text{DM}$ is derived according to Eq.~(\ref{j_dm}), assuming a spin parameter $\lambda=0.034$ \citep{Maccio}. For the baryonic components, the second column lists the measured values of $j_k$, while the third and fourth columns provide the values obtained by extrapolating the rotation curve out to $20$~kpc and $35$~kpc, respectively.}
\label{tab:j_k}
\centering
\begin{tabularx}{0.99\columnwidth}{l Y Y Y}
\toprule
\toprule
\multicolumn{4}{c}{Specific angular momentum results for the Big Wheel} \\
\midrule
\midrule
$j_k$ & Measured & Extrapolated (20 kpc) & Extrapolated (35 kpc)\tnote{a} \\
\midrule
$j_\star$ & $1594^{+70}_{-80}$ & $2356^{+26}_{-229}$ & $2575^{+341}_{-321}$ \\
$j_{\rm gas}$ & $1571^{+72}_{-81}$ & $2285^{+225}_{-227}$ & $2474^{+334}_{-310}$ \\
$j_{\rm bar}$ & $1592^{+62}_{-72}$ & $2353^{+194}_{-210}$ & $2572^{+306}_{-293}$ \\
\midrule
$j_{\rm DM}$ & - & \multicolumn{2}{c}{$1151^{+651}_{-262}$}\\
\bottomrule
\bottomrule
\end{tabularx}
\begin{tablenotes}
       \item [(a)] Convergence radius.
     \end{tablenotes}
  \end{threeparttable}
\end{table}

\subsection{Additional constraints on the baryonic component}
In addition, to put constraints on the mass of the DM halo associated with the Big Wheel, the dynamical model presented in this work provides additional information on the baryonic content of this galaxy.
In particular, we note that across multiple runs of the Bayesian analysis, with varying priors, constraints, and likelihood assumptions, the best-fit posterior for $M_\star$ resulted systematically lower (i.e., by a factor of 2) with respect to the central value in \cite{Galbiati25} estimated from fitting the SED, often reaching the minimum allowed by the prior. 
In this context, adopting an NFW halo requires a reduction of the stellar disc mass by approximately a factor of two with respect to the observational estimates. The large uncertainties affecting both stellar and gas mass estimates introduce significant freedom in the DM component, reflecting the well-known disc–halo degeneracy commonly encountered in galaxy mass modelling. We stress that, within our dynamical framework, fitting the rotation curve alone does not provide sufficiently strong constraints on the mass and distribution of DM in high-redshift galaxies as the Big Wheel. \citep[i.e., see][for more details]{Lelli23, RomanOliveira26}.
To investigate the possible origin of the deviation between photometry-based and dynamical measurements for the stellar mass, we performed some experiments by forcing the stellar mass to the SED-derived value of $M_\star=2.3\times10^{11} \msun$, and allowing the logarithmic baryon fraction to vary within increasingly wider prior ranges, extending the upper bound up to $10^{2}$, $10^{3}$, $10^{4}$, and $10^{10}$, while maintaining a lower limit of $10^{-4}$.
As a result, in all cases, the inferred DM content of the halo is systematically lower, reaching values down to $10^5 \ \msun$, and thus resulting in halo baryonic fraction as high as $\approx 10^5$, which is several orders of magnitude above the cosmological value. Given that these low values for the stellar and DM halo masses are consistent with the absence of a DM component, we performed a dynamical model excluding the DM halo. The results, presented in Appendix~\ref{appendix:nodmmodel}, accurately reproduces the observed rotation curve. Within the current cosmological framework, this suggests a clear tension between the SED-derived stellar mass and the dynamical estimate for the Big Wheel. In particular, under the assumption that the dynamical estimates provide a more accurate constraint on the total stellar mass, our result would imply that the photometry-based SED fitting overestimates the stellar mass by a factor of two.
One possible explanation is that some of the assumptions adopted in the stellar mass determination from SED-fitting see Section~\ref{sec:optical_data} may contribute to this inconsistency. For instance, the limited photometric information might prevent a proper constraint on the star formation history of the galaxy and on the age of the stellar population, leading to significant systematic uncertainties in the inferred stellar mass. Alternatively, the peculiar properties of the Big Wheel in terms, e.g. of SHMR and size as discussed above, might hint at the possibility that its star formation history could be very different with respect to the parametric shapes usually assumed by SED-fitting models.
However, W25 showed that including the spectrum and considering a non-parametric SFH does not help to resolve the discrepancy with the dynamical model. On the contrary, a non-parametric SFH results in a even higher $M_{\star}=3.7^{+2.6}_{-2.2}\times10^{11} \ \msun$, which is by itself inconsistent with the dynamical measurement without even considering additional dynamical components. The inclusion of the spectroscopic information available from JWST NIRSpec slit observations does not change this result since the detected continuum appears featureless in terms of absorption lines at current depth. 

Since our analysis is based on idealised models, several caveats should be taken into account. First of all, due to resolution limits, the measurements regarding the innermost part of both the Big Wheel and the ADF22.A1 galaxies are uncertain. As a consequence, the parameters related to the presence of a central bulge are less constrained and, in the case of a barred-like structure, they are difficult to be modelled, even if they can play a non-trivial role in the fit of the velocity curve and, finally, in the mass decomposition. In particular, the effective radius of the bulge component reflects the choice of the uniform distribution we imposed. 
Indeed the AGN is associated to a X-ray luminosity of $L_{2-10 \ \mathrm{keV}}, \approx10^{44} \mathrm{erg \ s^{-1}}$ \citep{Travascio25} consistent with a moderate-luminosity Seyfert galaxy. Under this assumption, even assuming a highly accreting black hole, this luminosity implies a central MBH mass that is of the order of $10^8 \ \msun$, which is at least two orders of magnitude smaller than the estimated mass of the bulge component. 

We stress that the available data obtained from CO(4--3) does not allow extending the rotation curve to radii much larger than $10$ kpc, which corresponds to a fraction of $\approx1/9$ of the virial radius of the associated DM halo derived from our dynamical analysis. Unfortunately, no other molecular or atomic tracers such as lower-J CO transitions or CII $158 \mu \rm m$ can be observed from the ground at the redshift of the Big Wheel. These tracers are typically associated with a more extended gas component than that traced by the CO(4--3) line.

\section{Summary}
\label{sec:conclusion}
In this work, we performed dynamical modelling to infer the dark matter halo mass and to provide further constraints on the baryonic components of the Big Wheel, the largest disc galaxy currently detected at $z>3$ (W25) whose origin is still unclear. In addition, with a numerical simulation of an idealised galaxy evolving adiabiatically, we confirmed the dynamical stability of the disc as derived from the analytical model. 

By comparing these analytical results to deep ALMA CO(4--3) observations (Section~\ref{sec:almaC0}), we derived the following properties for the Big Wheel:
\begin{itemize}
    \item a dark matter (DM) halo mass of $\log(M_h/\msun)=12.11^{+0.29}_{-0.17}$ and a dynamical stellar mass of $\log(M_\star/\msun)=11.00^{+0.11}_{-0.12}$ obtained by using as priors the SED fitting results of \cite{Galbiati25} and the concentration-mass relation by \cite{Dutton14}. The resulting SHM ratio of the Big Wheel is thus $M_\star/M_h=0.06^{+0.04}_{-0.03}$. In addition, we obtained a dynamical gas mass of $\log(M_\mathrm{gas}/\msun)=10.76^{+0.13}_{-0.12}$ and a bulge mass of $\log(M_\mathrm{bulge}/\msun)=9.93^{+0.19}_{-0.23}$, resulting in a gas fraction of approximately $0.6$ and a bulge to disc ratio of $B/T\approx0.09$. However, the latter parameters are less constrained due to uncertainties in the conversion factor between CO and total gas mass.
    \item the stellar-to-halo mass (SHM) ratio of the Big Wheel is $\approx3$ times higher than expected for the general galaxy population at $z\sim3$ given current halo occupation distribution (HOD)-based models (e.g., \citealt{Moster13} and \citealt{Shuntov22}). Instead, the Big Wheel SHM ratio is similar to the value estimated for local super-spiral galaxies, a very rare population of local massive disc galaxies in which star formation has not quenched \citep{Ogle16, Ogle19, DiTeodoro21, DiTeodoro23}. Our result suggest that the Big Wheel has been able to efficiently assemble its large stellar mass without experiencing strong feedback episodes. 
    \item by assuming a value for the ratio of the specific angular momentum of the Big Wheel and its parent halo ($\mathcal{R}_j$), our results allow us to obtain an estimate of the halo spin parameter $\lambda$ following the formalism introduced by \citep{Mo98} as discussed in Section~\ref{physical_implications}. 
   In particular, by assuming $\mathcal{R}_j=1$ \citep{Mo98, Somerville99, Cole00} or $\mathcal{R}_j=0.61$ \citep{Dutton12}, we obtain a spin parameter for the halo of the Big Wheel of $\lambda_\text{BW}=0.085$ or $\lambda_\text{BW}=0.139$, respectively. These values are approximately two and four times larger than the expected value for a typical halo as calculated by \cite{Bullock01}. In the context of the disc formation model suggested by \citep{Mo98}, such large $\lambda$  would explain the fact that the Big Wheel disc size is about three times larger than expected from the size-mass relation at its redshift as discussed in W25. Performing a complementary analysis and fixing the spin parameter at the peak value of its distribution ($\lambda=0.034$) as in \cite{DiTeodoro23}, we find a stellar-to-halo specific angular momentum ratio of $f_{j_{\star, \rm BW}}=2.06^{+0.61}_{-0.71}$, and the one for the total baryonic content equal to $f_{j_{\rm bar, BW}}=2.05^{+0.59}_{-0.71}$. Applying the same procedures to the ADF22.A1 galaxy, we find $\lambda_\text{ADF22.A1}=0.038$ or $\lambda_\text{ADF22.A1}=0.063$. As for the specific angular momentum of the galactic components, normalized to that of the DM, they are equal to $f_{j_{\star, \rm ADF22.A1}}=0.92^{+0.76}_{-0.43}$ and $f_{j_{\rm bar, ADF22.A1}}=0.94^{+0.76}_{-0.44}$. These findings, in agreement with \cite{Rizzo26}, suggest that the large spatial extent of these galaxies may originate from different physical drivers: the high mass of the host halo in ADF22.A1, and a particularly high halo spin in the Big Wheel. However, other explanations cannot be excluded at the moment.
    \item by performing numerical simulations of an idealized galaxy evolved in isolation and adiabatically (Section~\ref{sec:num_sim}), we confirmed that a galaxy with physical properties as found by our model is stable against gravitational instabilities for at least $2.5$ Gyr, which corresponds to $15$ dynamical times. Moreover, we observed the development of spiral arms and the presence of a central pseudo-bulge.
\end{itemize}

Despite the simplicity of our model and the observational limitations, our results suggest that galaxies such as the Big Wheel can be used as a unique laboratory to test our understanding of gas feeding and feedback at high-redshift. 
Further efforts are needed to understand, from an observational point of view, how rare are massive disc galaxies such as the Big Wheel at $z\gtrsim3$ and what the connection between these systems and Cosmic Web nodes is. Given the paucity of surveys targeting overdense regions such as the MQN01 structure hosting the Big Wheel, new wide-field surveys including high-resolution Infrared imaging would be necessary. On the modelling side, further efforts are needed to understand the physical origin and evolutionary path of such systems by including them into their cosmological context.

\begin{acknowledgements}
This project was supported by the European Research Council (ERC) Consolidator grant no. 864361 (CosmicWeb) and by Fondazione Cariplo grant no. 2020-0902. GQ acknowledges support from the FARE2020 CosmicNodes grant no. R20S99FS3J) project. AP acknowledges support from the Independent Research Fund Denmark (DFF) under grant no. 3120-00043B. CB acknowledges support from the Carlsberg Foundation Fellowship Programme by Carlsbergfondet. This paper makes use of the following ALMA data: ADS/JAO.ALMA$\#$2021.1.00793.S,$\#$2025.1.00107.S. 
This work is based in part on observations made with the NASA/ESA/CSA James Webb Space Telescope. 
These observations are associated with program $\#$1835. Support for program $\#$1835 was provided by NASA through a grant from the Space Telescope Science Institute, which is operated by the Association of Universities for Research in Astronomy, Inc., under NASA contract NAS 5-03127.
\end{acknowledgements} 

\bibliographystyle{aa}
\bibliography{./Biblio}

\begin{thebibliography}{75}
\expandafter\ifx\csname natexlab\endcsname\relax\def\natexlab#1{#1}\fi

\bibitem[{{Behroozi} {et~al.}(2019){Behroozi}, {Wechsler}, {Hearin}, \& {Conroy}}]{Behroozi19}
{Behroozi}, P., {Wechsler}, R.~H., {Hearin}, A.~P., \& {Conroy}, C. 2019, \mnras, 488, 3143

\bibitem[{{Behroozi} {et~al.}(2013){Behroozi}, {Wechsler}, \& {Conroy}}]{Behroozi13}
{Behroozi}, P.~S., {Wechsler}, R.~H., \& {Conroy}, C. 2013, \apj, 770, 57

\bibitem[{{Bolatto} {et~al.}(2013){Bolatto}, {Wolfire}, \& {Leroy}}]{Bolatto13}
{Bolatto}, A.~D., {Wolfire}, M., \& {Leroy}, A.~K. 2013, \araa, 51, 207

\bibitem[{{Bonoli} {et~al.}(2016){Bonoli}, {Mayer}, {Kazantzidis}, {Madau}, {Bellovary}, \& {Governato}}]{Bonoli16}
{Bonoli}, S., {Mayer}, L., {Kazantzidis}, S., {et~al.} 2016, \mnras, 459, 2603

\bibitem[{{Boquien} {et~al.}(2019){Boquien}, {Burgarella}, {Roehlly}, {Buat}, {Ciesla}, {Corre}, {Inoue}, \& {Salas}}]{Boquien19}
{Boquien}, M., {Burgarella}, D., {Roehlly}, Y., {et~al.} 2019, \aap, 622, A103

\bibitem[{{Bruzual} \& {Charlot}(2003)}]{Bruzual03}
{Bruzual}, G. \& {Charlot}, S. 2003, \mnras, 344, 1000

\bibitem[{{Bullock} {et~al.}(2001){Bullock}, {Dekel}, {Kolatt}, {Kravtsov}, {Klypin}, {Porciani}, \& {Primack}}]{Bullock01}
{Bullock}, J.~S., {Dekel}, A., {Kolatt}, T.~S., {et~al.} 2001, \apj, 555, 240

\bibitem[{{Burgarella} {et~al.}(2005){Burgarella}, {Buat}, \& {Iglesias-P{\'a}ramo}}]{Burganella05}
{Burgarella}, D., {Buat}, V., \& {Iglesias-P{\'a}ramo}, J. 2005, \mnras, 360, 1413

\bibitem[{{Calzetti} {et~al.}(2000){Calzetti}, {Armus}, {Bohlin}, {Kinney}, {Koornneef}, \& {Storchi-Bergmann}}]{Calzetti00}
{Calzetti}, D., {Armus}, L., {Bohlin}, R.~C., {et~al.} 2000, \apj, 533, 682

\bibitem[{{Carilli} \& {Walter}(2013)}]{Carilli13}
{Carilli}, C.~L. \& {Walter}, F. 2013, \araa, 51, 105

\bibitem[{{Chabrier}(2003)}]{Chabrier03}
{Chabrier}, G. 2003, \pasp, 115, 763

\bibitem[{{Cole} {et~al.}(2000){Cole}, {Lacey}, {Baugh}, \& {Frenk}}]{Cole00}
{Cole}, S., {Lacey}, C.~G., {Baugh}, C.~M., \& {Frenk}, C.~S. 2000, \mnras, 319, 168

\bibitem[{{Croton} {et~al.}(2006){Croton}, {Springel}, {White}, {De Lucia}, {Frenk}, {Gao}, {Jenkins}, {Kauffmann}, {Navarro}, \& {Yoshida}}]{Croton06}
{Croton}, D.~J., {Springel}, V., {White}, S. D.~M., {et~al.} 2006, \mnras, 365, 11

\bibitem[{{Dekel} \& {Silk}(1986)}]{Dekel86}
{Dekel}, A. \& {Silk}, J. 1986, \apj, 303, 39

\bibitem[{{Di Teodoro} \& {Fraternali}(2015)}]{2015DiTeodoro}
{Di Teodoro}, E.~M. \& {Fraternali}, F. 2015, \mnras, 451, 3021

\bibitem[{{Di Teodoro} {et~al.}(2023){Di Teodoro}, {Posti}, {Fall}, {Ogle}, {Jarrett}, {Appleton}, {Cluver}, {Haynes}, \& {Lisenfeld}}]{DiTeodoro23}
{Di Teodoro}, E.~M., {Posti}, L., {Fall}, S.~M., {et~al.} 2023, \mnras, 518, 6340

\bibitem[{{Di Teodoro} {et~al.}(2021){Di Teodoro}, {Posti}, {Ogle}, {Fall}, \& {Jarrett}}]{DiTeodoro21}
{Di Teodoro}, E.~M., {Posti}, L., {Ogle}, P.~M., {Fall}, S.~M., \& {Jarrett}, T. 2021, \mnras, 507, 5820

\bibitem[{{Dutton} \& {Macci{\`o}}(2014)}]{Dutton14}
{Dutton}, A.~A. \& {Macci{\`o}}, A.~V. 2014, \mnras, 441, 3359

\bibitem[{{Dutton} \& {van den Bosch}(2012)}]{Dutton12}
{Dutton}, A.~A. \& {van den Bosch}, F.~C. 2012, \mnras, 421, 608

\bibitem[{{Faucher-Gigu{\`e}re} {et~al.}(2011){Faucher-Gigu{\`e}re}, {Kere{\v{s}}}, \& {Ma}}]{FaucherGiguere11}
{Faucher-Gigu{\`e}re}, C.-A., {Kere{\v{s}}}, D., \& {Ma}, C.-P. 2011, \mnras, 417, 2982

\bibitem[{{Gadotti}(2009)}]{Gadotti09}
{Gadotti}, D.~A. 2009, \mnras, 393, 1531

\bibitem[{{Galbiati} {et~al.}(2025){Galbiati}, {Cantalupo}, {Steidel}, {Pensabene}, {Travascio}, {Wang}, {Fossati}, {Fumagalli}, {Rudie}, {Fresco}, {Lazeyras}, {Ledos}, \& {Quadri}}]{Galbiati25}
{Galbiati}, M., {Cantalupo}, S., {Steidel}, C., {et~al.} 2025, \aap, 696, A95

\bibitem[{{Girelli} {et~al.}(2020){Girelli}, {Pozzetti}, {Bolzonella}, {Giocoli}, {Marulli}, \& {Baldi}}]{Girelli20}
{Girelli}, G., {Pozzetti}, L., {Bolzonella}, M., {et~al.} 2020, \aap, 634, A135

\bibitem[{{Hopkins} {et~al.}(2014){Hopkins}, {Kere{\v{s}}}, {O{\~n}orbe}, {Faucher-Gigu{\`e}re}, {Quataert}, {Murray}, \& {Bullock}}]{Hopkins14}
{Hopkins}, P.~F., {Kere{\v{s}}}, D., {O{\~n}orbe}, J., {et~al.} 2014, \mnras, 445, 581

\bibitem[{{Jiang} {et~al.}(2025){Jiang}, {Liang}, {Jin}, {Gao}, {Wang}, {Cantalupo}, {Shen}, {Ho}, {Peng}, \& {Wang}}]{Fangzhou25}
{Jiang}, F., {Liang}, J., {Jin}, B., {et~al.} 2025, arXiv e-prints, arXiv:2504.01070

\bibitem[{{Kormendy} \& {Kennicutt}(2004)}]{Kormendy04}
{Kormendy}, J. \& {Kennicutt}, Jr., R.~C. 2004, \araa, 42, 603

\bibitem[{{Kravtsov}(2013)}]{Kravtsov13}
{Kravtsov}, A.~V. 2013, \apjl, 764, L31

\bibitem[{{Lelli} {et~al.}(2014){Lelli}, {Verheijen}, \& {Fraternali}}]{2014Lelli}
{Lelli}, F., {Verheijen}, M., \& {Fraternali}, F. 2014, \mnras, 445, 1694

\bibitem[{{Lelli} {et~al.}(2023){Lelli}, {Zhang}, {Bisbas}, {Lin}, {Papadopoulos}, {Schombert}, {Di Teodoro}, {Marasco}, \& {McGaugh}}]{Lelli23}
{Lelli}, F., {Zhang}, Z.-Y., {Bisbas}, T.~G., {et~al.} 2023, \aap, 672, A106

\bibitem[{{Lima Neto} {et~al.}(1999){Lima Neto}, {Gerbal}, \& {M{\'a}rquez}}]{LimaNeto99}
{Lima Neto}, G.~B., {Gerbal}, D., \& {M{\'a}rquez}, I. 1999, \mnras, 309, 481

\bibitem[{{Mac Low} \& {Ferrara}(1999)}]{MacLow99}
{Mac Low}, M.-M. \& {Ferrara}, A. 1999, \apj, 513, 142

\bibitem[{{Macci{\`o}} {et~al.}(2007){Macci{\`o}}, {Dutton}, {van den Bosch}, {Moore}, {Potter}, \& {Stadel}}]{Maccio}
{Macci{\`o}}, A.~V., {Dutton}, A.~A., {van den Bosch}, F.~C., {et~al.} 2007, \mnras, 378, 55

\bibitem[{{Marasco} {et~al.}(2023){Marasco}, {Belfiore}, {Cresci}, {Lelli}, {Venturi}, {Hunt}, {Concas}, {Marconi}, {Mannucci}, {Mingozzi}, {McLeod}, {Kumari}, {Carniani}, {Vanzi}, \& {Ginolfi}}]{2023Marasco}
{Marasco}, A., {Belfiore}, F., {Cresci}, G., {et~al.} 2023, \aap, 670, A92

\bibitem[{{McMullin} {et~al.}(2007){McMullin}, {Waters}, {Schiebel}, {Young}, \& {Golap}}]{McMullin07}
{McMullin}, J.~P., {Waters}, B., {Schiebel}, D., {Young}, W., \& {Golap}, K. 2007, in Astronomical Society of the Pacific Conference Series, Vol. 376, Astronomical Data Analysis Software and Systems XVI, ed. R.~A. {Shaw}, F.~{Hill}, \& D.~J. {Bell}, 127

\bibitem[{{Mo} {et~al.}(1998){Mo}, {Mao}, \& {White}}]{Mo98}
{Mo}, H.~J., {Mao}, S., \& {White}, S. D.~M. 1998, \mnras, 295, 319

\bibitem[{{Moster} {et~al.}(2013){Moster}, {Naab}, \& {White}}]{Moster13}
{Moster}, B.~P., {Naab}, T., \& {White}, S. D.~M. 2013, \mnras, 428, 3121

\bibitem[{{Moster} {et~al.}(2018){Moster}, {Naab}, \& {White}}]{Moster18}
{Moster}, B.~P., {Naab}, T., \& {White}, S. D.~M. 2018, \mnras, 477, 1822

\bibitem[{{Navarro} {et~al.}(1996){Navarro}, {Frenk}, \& {White}}]{Navarro96}
{Navarro}, J.~F., {Frenk}, C.~S., \& {White}, S. D.~M. 1996, \apj, 462, 563

\bibitem[{{Noll} {et~al.}(2009){Noll}, {Burgarella}, {Giovannoli}, {Buat}, {Marcillac}, \& {Mu{\~n}oz-Mateos}}]{Noll09}
{Noll}, S., {Burgarella}, D., {Giovannoli}, E., {et~al.} 2009, \aap, 507, 1793

\bibitem[{{Ogle} {et~al.}(2019){Ogle}, {Lanz}, {Appleton}, {Helou}, \& {Mazzarella}}]{Ogle19}
{Ogle}, P.~M., {Lanz}, L., {Appleton}, P.~N., {Helou}, G., \& {Mazzarella}, J. 2019, \apjs, 243, 14

\bibitem[{{Ogle} {et~al.}(2016){Ogle}, {Lanz}, {Nader}, \& {Helou}}]{Ogle16}
{Ogle}, P.~M., {Lanz}, L., {Nader}, C., \& {Helou}, G. 2016, \apj, 817, 109

\bibitem[{{Ormerod} {et~al.}(2024){Ormerod}, {Conselice}, {Adams}, {Harvey}, {Austin}, {Trussler}, {Ferreira}, {Caruana}, {Lucatelli}, {Li}, \& {Roper}}]{Ormerod24}
{Ormerod}, K., {Conselice}, C.~J., {Adams}, N.~J., {et~al.} 2024, \mnras, 527, 6110

\bibitem[{{Paquereau} {et~al.}(2025){Paquereau}, {Laigle}, {McCracken}, {Shuntov}, {Ilbert}, {Akins}, {Allen}, {Arango-Togo}, {Berman}, {B{\'e}thermin}, {Casey}, {McCleary}, {Dubois}, {Drakos}, {Faisst}, {Franco}, {Harish}, {Jespersen}, {Kartaltepe}, {Koekemoer}, {Kokorev}, {Lambrides}, {Larson}, {Liu}, {Le Borgne}, {Lewis}, {McKinney}, {Mercier}, {Rhodes}, {Robertson}, {Toft}, {Trebitsch}, {Tresse}, \& {Weaver}}]{Paquereau25}
{Paquereau}, L., {Laigle}, C., {McCracken}, H.~J., {et~al.} 2025, \aap, 702, A163

\bibitem[{{Pensabene} {et~al.}(2024){Pensabene}, {Cantalupo}, {Cicone}, {Decarli}, {Galbiati}, {Ginolfi}, {de Beer}, {Fossati}, {Fumagalli}, {Lazeyras}, {Pezzulli}, {Travascio}, {Wang}, {Matthee}, \& {Maseda}}]{Pensabene24}
{Pensabene}, A., {Cantalupo}, S., {Cicone}, C., {et~al.} 2024, \aap, 684, A119

\bibitem[{{Pensabene} {et~al.}(2025){Pensabene}, {Cantalupo}, {Wang}, {Bacchini}, {Fraternali}, {Bischetti}, {Cicone}, {Decarli}, {Pezzulli}, {Galbiati}, {Lazeyras}, {Ledos}, {Quadri}, \& {Travascio}}]{Pensabene25}
{Pensabene}, A., {Cantalupo}, S., {Wang}, W., {et~al.} 2025, \aap, 701, A120

\bibitem[{{Perret} {et~al.}(2014){Perret}, {Renaud}, {Epinat}, {Amram}, {Bournaud}, {Contini}, {Teyssier}, \& {Lambert}}]{Perret14}
{Perret}, V., {Renaud}, F., {Epinat}, B., {et~al.} 2014, \aap, 562, A1

\bibitem[{{Planck Collaboration} {et~al.}(2020){Planck Collaboration}, {Aghanim}, {Akrami}, {Ashdown}, {Aumont}, {Baccigalupi}, {Ballardini}, {Banday}, {Barreiro}, {Bartolo}, {Basak}, {Battye}, {Benabed}, {Bernard}, {Bersanelli}, {Bielewicz}, {Bock}, {Bond}, {Borrill}, {Bouchet}, {Boulanger}, {Bucher}, {Burigana}, {Butler}, {Calabrese}, {Cardoso}, {Carron}, {Challinor}, {Chiang}, {Chluba}, {Colombo}, {Combet}, {Contreras}, {Crill}, {Cuttaia}, {de Bernardis}, {de Zotti}, {Delabrouille}, {Delouis}, {Di Valentino}, {Diego}, {Dor{\'e}}, {Douspis}, {Ducout}, {Dupac}, {Dusini}, {Efstathiou}, {Elsner}, {En{\ss}lin}, {Eriksen}, {Fantaye}, {Farhang}, {Fergusson}, {Fernandez-Cobos}, {Finelli}, {Forastieri}, {Frailis}, {Fraisse}, {Franceschi}, {Frolov}, {Galeotta}, {Galli}, {Ganga}, {G{\'e}nova-Santos}, {Gerbino}, {Ghosh}, {Gonz{\'a}lez-Nuevo}, {G{\'o}rski}, {Gratton}, {Gruppuso}, {Gudmundsson}, {Hamann}, {Handley}, {Hansen}, {Herranz}, {Hildebrandt}, {Hivon}, {Huang}, {Jaffe}, {Jones}, {Karakci}, {Keih{\"a}nen},
  {Keskitalo}, {Kiiveri}, {Kim}, {Kisner}, {Knox}, {Krachmalnicoff}, {Kunz}, {Kurki-Suonio}, {Lagache}, {Lamarre}, {Lasenby}, {Lattanzi}, {Lawrence}, {Le Jeune}, {Lemos}, {Lesgourgues}, {Levrier}, {Lewis}, {Liguori}, {Lilje}, {Lilley}, {Lindholm}, {L{\'o}pez-Caniego}, {Lubin}, {Ma}, {Mac{\'\i}as-P{\'e}rez}, {Maggio}, {Maino}, {Mandolesi}, {Mangilli}, {Marcos-Caballero}, {Maris}, {Martin}, {Martinelli}, {Mart{\'\i}nez-Gonz{\'a}lez}, {Matarrese}, {Mauri}, {McEwen}, {Meinhold}, {Melchiorri}, {Mennella}, {Migliaccio}, {Millea}, {Mitra}, {Miville-Desch{\^e}nes}, {Molinari}, {Montier}, {Morgante}, {Moss}, {Natoli}, {N{\o}rgaard-Nielsen}, {Pagano}, {Paoletti}, {Partridge}, {Patanchon}, {Peiris}, {Perrotta}, {Pettorino}, {Piacentini}, {Polastri}, {Polenta}, {Puget}, {Rachen}, {Reinecke}, {Remazeilles}, {Renzi}, {Rocha}, {Rosset}, {Roudier}, {Rubi{\~n}o-Mart{\'\i}n}, {Ruiz-Granados}, {Salvati}, {Sandri}, {Savelainen}, {Scott}, {Shellard}, {Sirignano}, {Sirri}, {Spencer}, {Sunyaev}, {Suur-Uski}, {Tauber}, {Tavagnacco},
  {Tenti}, {Toffolatti}, {Tomasi}, {Trombetti}, {Valenziano}, {Valiviita}, {Van Tent}, {Vibert}, {Vielva}, {Villa}, {Vittorio}, {Wandelt}, {Wehus}, {White}, {White}, {Zacchei}, \& {Zonca}}]{Planck20}
{Planck Collaboration}, {Aghanim}, N., {Akrami}, Y., {et~al.} 2020, \aap, 641, A6

\bibitem[{{Posti} \& {Fall}(2021)}]{Posti21}
{Posti}, L. \& {Fall}, S.~M. 2021, \aap, 649, A119

\bibitem[{{Prugniel} \& {Simien}(1997)}]{Prugniel97}
{Prugniel}, P. \& {Simien}, F. 1997, \aap, 321, 111

\bibitem[{{Rizzo} {et~al.}(2026){Rizzo}, {Mancera Pi{\~n}a}, {Pezzulli}, \& {Despali}}]{Rizzo26}
{Rizzo}, F., {Mancera Pi{\~n}a}, P.~E., {Pezzulli}, G., \& {Despali}, G. 2026, arXiv e-prints, arXiv:2604.07440

\bibitem[{{Roman-Oliveira} {et~al.}(2023){Roman-Oliveira}, {Fraternali}, \& {Rizzo}}]{2023RomanOliveira}
{Roman-Oliveira}, F., {Fraternali}, F., \& {Rizzo}, F. 2023, \mnras, 521, 1045

\bibitem[{{Roman-Oliveira} {et~al.}(2024){Roman-Oliveira}, {Rizzo}, \& {Fraternali}}]{RomanOliveira24}
{Roman-Oliveira}, F., {Rizzo}, F., \& {Fraternali}, F. 2024, \aap, 687, A35

\bibitem[{{Roman-Oliveira} {et~al.}(2026){Roman-Oliveira}, {Rizzo}, \& {Fraternali}}]{RomanOliveira26}
{Roman-Oliveira}, F., {Rizzo}, F., \& {Fraternali}, F. 2026, \mnras [\eprint[arXiv]{2601.03338}]

\bibitem[{{Shen} {et~al.}(2003){Shen}, {Mo}, {White}, {Blanton}, {Kauffmann}, {Voges}, {Brinkmann}, \& {Csabai}}]{Shen03}
{Shen}, S., {Mo}, H.~J., {White}, S. D.~M., {et~al.} 2003, \mnras, 343, 978

\bibitem[{{Shuntov} {et~al.}(2022){Shuntov}, {McCracken}, {Gavazzi}, {Laigle}, {Weaver}, {Davidzon}, {Ilbert}, {Kauffmann}, {Faisst}, {Dubois}, {Koekemoer}, {Moneti}, {Milvang-Jensen}, {Mobasher}, {Sanders}, \& {Toft}}]{Shuntov22}
{Shuntov}, M., {McCracken}, H.~J., {Gavazzi}, R., {et~al.} 2022, \aap, 664, A61

\bibitem[{{Silk} \& {Mamon}(2012)}]{silk12}
{Silk}, J. \& {Mamon}, G.~A. 2012, Research in Astronomy and Astrophysics, 12, 917

\bibitem[{{Somerville} {et~al.}(2018){Somerville}, {Behroozi}, {Pandya}, {Dekel}, {Faber}, {Fontana}, {Koekemoer}, {Koo}, {P{\'e}rez-Gonz{\'a}lez}, {Primack}, {Santini}, {Taylor}, \& {van der Wel}}]{Somerville18}
{Somerville}, R.~S., {Behroozi}, P., {Pandya}, V., {et~al.} 2018, \mnras, 473, 2714

\bibitem[{{Somerville} \& {Dav{\'e}}(2015)}]{Somerville15}
{Somerville}, R.~S. \& {Dav{\'e}}, R. 2015, \araa, 53, 51

\bibitem[{{Somerville} \& {Primack}(1999)}]{Somerville99}
{Somerville}, R.~S. \& {Primack}, J.~R. 1999, \mnras, 310, 1087

\bibitem[{{Speagle}(2020)}]{Speagle20}
{Speagle}, J.~S. 2020, \mnras, 493, 3132

\bibitem[{{Terzi{\'c}} \& {Graham}(2005)}]{Terzic05}
{Terzi{\'c}}, B. \& {Graham}, A.~W. 2005, \mnras, 362, 197

\bibitem[{{Teyssier}(2002)}]{Teyssier02}
{Teyssier}, R. 2002, \aap, 385, 337

\bibitem[{{Travascio} {et~al.}(2025{\natexlab{a}}){Travascio}, {Cantalupo}, {Pezzulli}, {Tozzi}, {Di Mascolo}, {Esposito}, {Lazeyras}, {Lepore}, {Borgani}, {Elvis}, {Fabbiano}, {Galbiati}, {Ledos}, {Middei}, {Pensabene}, {Piconcelli}, {Quadri}, {Vito}, {Wang}, \& {Zappacosta}}]{Travascio25b}
{Travascio}, A., {Cantalupo}, S., {Pezzulli}, G., {et~al.} 2025{\natexlab{a}}, arXiv e-prints, arXiv:2508.20074

\bibitem[{{Travascio} {et~al.}(2025{\natexlab{b}}){Travascio}, {Cantalupo}, {Tozzi}, {Vito}, {Pezzulli}, {Paggi}, {Elvis}, {Fabbiano}, {Fiore}, {Fossati}, {Fresco}, {Fumagalli}, {Galbiati}, {Lazeyras}, {Ledos}, {Pannella}, {Pensabene}, {Quadri}, \& {Wang}}]{Travascio25}
{Travascio}, A., {Cantalupo}, S., {Tozzi}, P., {et~al.} 2025{\natexlab{b}}, \aap, 694, A165

\bibitem[{{Umehata} {et~al.}(2025){Umehata}, {Steidel}, {Smail}, {Swinbank}, {Monson}, {Rosario}, {Lehmer}, {Nakanishi}, {Kubo}, {Iono}, {Alexander}, {Kohno}, {Tamura}, {Ivison}, {Saito}, {Mitsuhashi}, {Huang}, \& {Matsuda}}]{Umehata25}
{Umehata}, H., {Steidel}, C.~C., {Smail}, I., {et~al.} 2025, \pasj, 77, 432

\bibitem[{{van Albada} {et~al.}(1985){van Albada}, {Bahcall}, {Begeman}, \& {Sancisi}}]{vanAlbada85}
{van Albada}, T.~S., {Bahcall}, J.~N., {Begeman}, K., \& {Sancisi}, R. 1985, \apj, 295, 305

\bibitem[{{van Asselt} {et~al.}(2026){van Asselt}, {Rizzo}, \& {Di Mascolo}}]{2026vanAsselt}
{van Asselt}, M., {Rizzo}, F., \& {Di Mascolo}, L. 2026, arXiv e-prints, arXiv:2601.03339

\bibitem[{{van der Wel} {et~al.}(2014){van der Wel}, {Franx}, {van Dokkum}, {Skelton}, {Momcheva}, {Whitaker}, {Brammer}, {Bell}, {Rix}, {Wuyts}, {Ferguson}, {Holden}, {Barro}, {Koekemoer}, {Chang}, {McGrath}, {H{\"a}ussler}, {Dekel}, {Behroozi}, {Fumagalli}, {Leja}, {Lundgren}, {Maseda}, {Nelson}, {Wake}, {Patel}, {Labb{\'e}}, {Faber}, {Grogin}, \& {Kocevski}}]{vanderWel14}
{van der Wel}, A., {Franx}, M., {van Dokkum}, P.~G., {et~al.} 2014, \apj, 788, 28

\bibitem[{{Verheijen} \& {Sancisi}(2001)}]{2001Verheijen}
{Verheijen}, M.~A.~W. \& {Sancisi}, R. 2001, \aap, 370, 765

\bibitem[{{Wang} {et~al.}(2025{\natexlab{a}}){Wang}, {Cantalupo}, {Pensabene}, {Galbiati}, {Travascio}, {Steidel}, {Maseda}, {Pezzulli}, {de Beer}, {Fossati}, {Fumagalli}, {Gallego}, {Lazeyras}, {Mackenzie}, {Matthee}, {Nanayakkara}, \& {Quadri}}]{Wang25}
{Wang}, W., {Cantalupo}, S., {Pensabene}, A., {et~al.} 2025{\natexlab{a}}, Nature Astronomy, 9, 710

\bibitem[{{Wang} {et~al.}(2025{\natexlab{b}}){Wang}, {Cantalupo}, {Wang}, {Galbiati}, {Steidel}, {Pensabene}, {Mao}, {Travascio}, {Lazeyras}, {Ledos}, \& {Quadri}}]{WangX25}
{Wang}, X., {Cantalupo}, S., {Wang}, W., {et~al.} 2025{\natexlab{b}}, arXiv e-prints, arXiv:2511.19608

\bibitem[{{Wechsler} \& {Tinker}(2018)}]{Wechsler18}
{Wechsler}, R.~H. \& {Tinker}, J.~L. 2018, \araa, 56, 435

\bibitem[{{White} \& {Rees}(1978)}]{White78}
{White}, S.~D.~M. \& {Rees}, M.~J. 1978, \mnras, 183, 341

\bibitem[{{Yang} {et~al.}(2022){Yang}, {Boquien}, {Brandt}, {Buat}, {Burgarella}, {Ciesla}, {Lehmer}, {Ma{\l}ek}, {Mountrichas}, {Papovich}, {Pons}, {Stalevski}, {Theul{\'e}}, \& {Zhu}}]{Yang22}
{Yang}, G., {Boquien}, M., {Brandt}, W.~N., {et~al.} 2022, \apj, 927, 192

\bibitem[{{Yang} {et~al.}(2020){Yang}, {Boquien}, {Buat}, {Burgarella}, {Ciesla}, {Duras}, {Stalevski}, {Brandt}, \& {Papovich}}]{Yang20}
{Yang}, G., {Boquien}, M., {Buat}, V., {et~al.} 2020, \mnras, 491, 740

\end{thebibliography}

\begin{appendix}
\section{Kinematic modelling}
\label{appendix:kinematic_mod}
\subsection{Disc geometry from JWST imaging}\label{sec:jwst_phot}
\begin{figure*}[h!]
    \centering
    \includegraphics[width=0.23\textwidth]{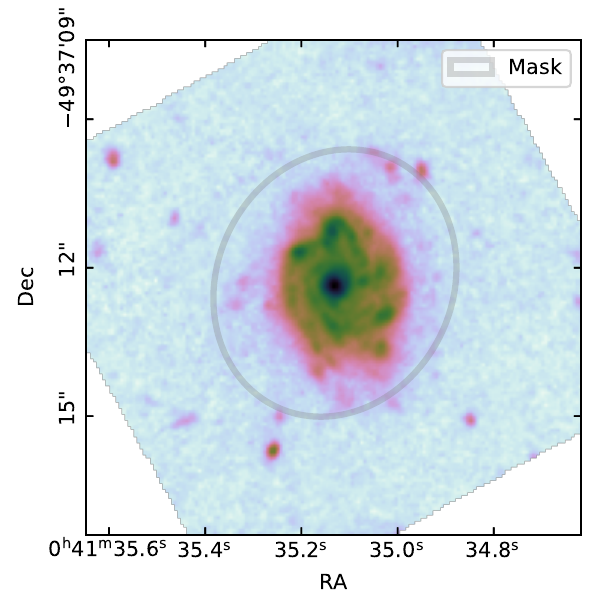}
    \includegraphics[width=0.23\textwidth]{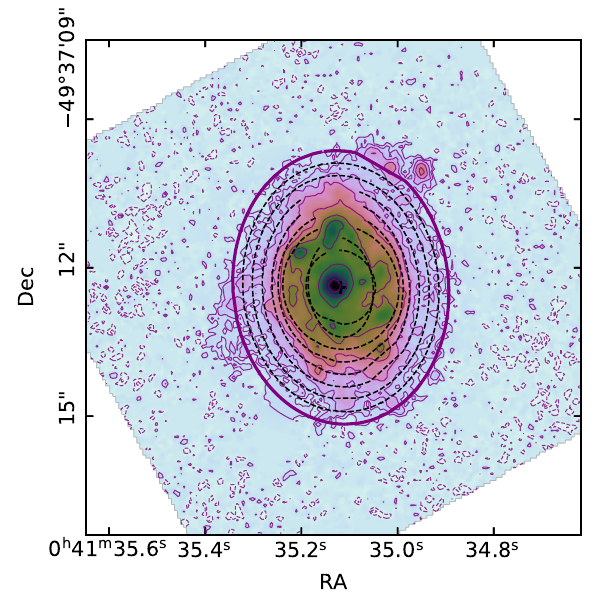}
\includegraphics[width=0.27\textwidth]{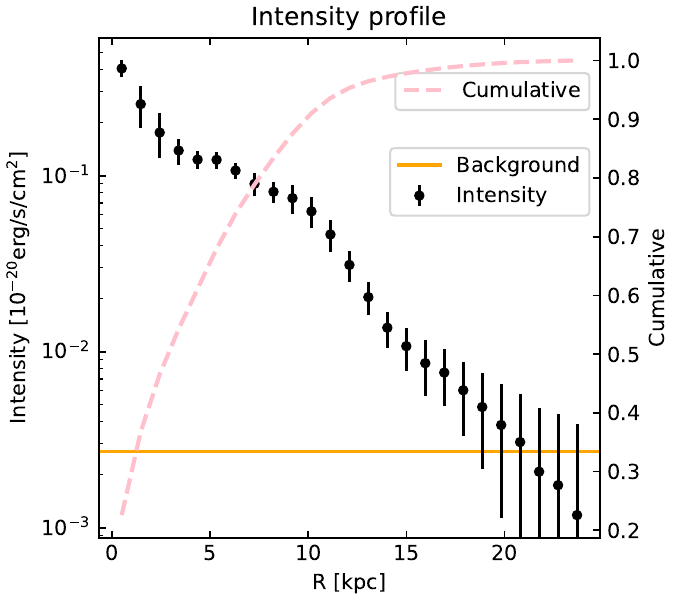}
    \caption{Photometry analysis. 
    \textit{First panel:} F322W2 image, with the red ellipse showing the mask applied to the image for the background calculation. 
    \textit{Second panel:} F322W2 image after background subtraction and source masking. The fitted isophote are  shown by the purples ellipses, with the thick one being the outermost. 
    \textit{Third panel:} Intensity profile (black points) and cumulative light distribution (pink curve). }
    \label{fig:jwst}
\end{figure*}

\begin{figure*}
 \centering
  \includegraphics[width=0.195\textwidth]{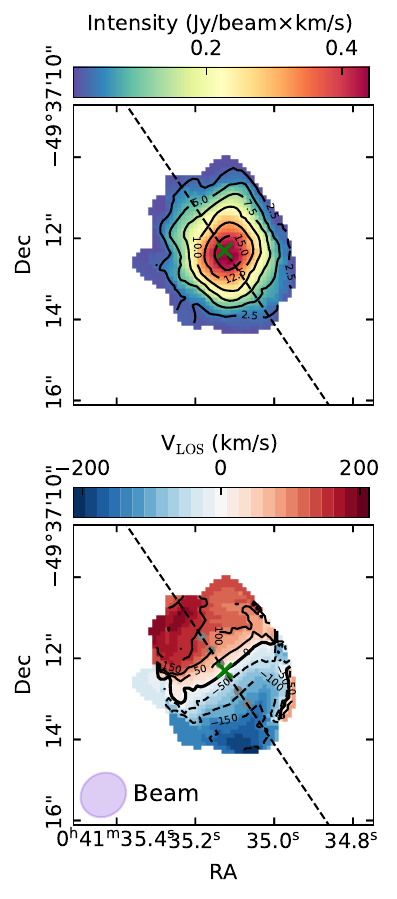}
\includegraphics[width=0.335\textwidth]{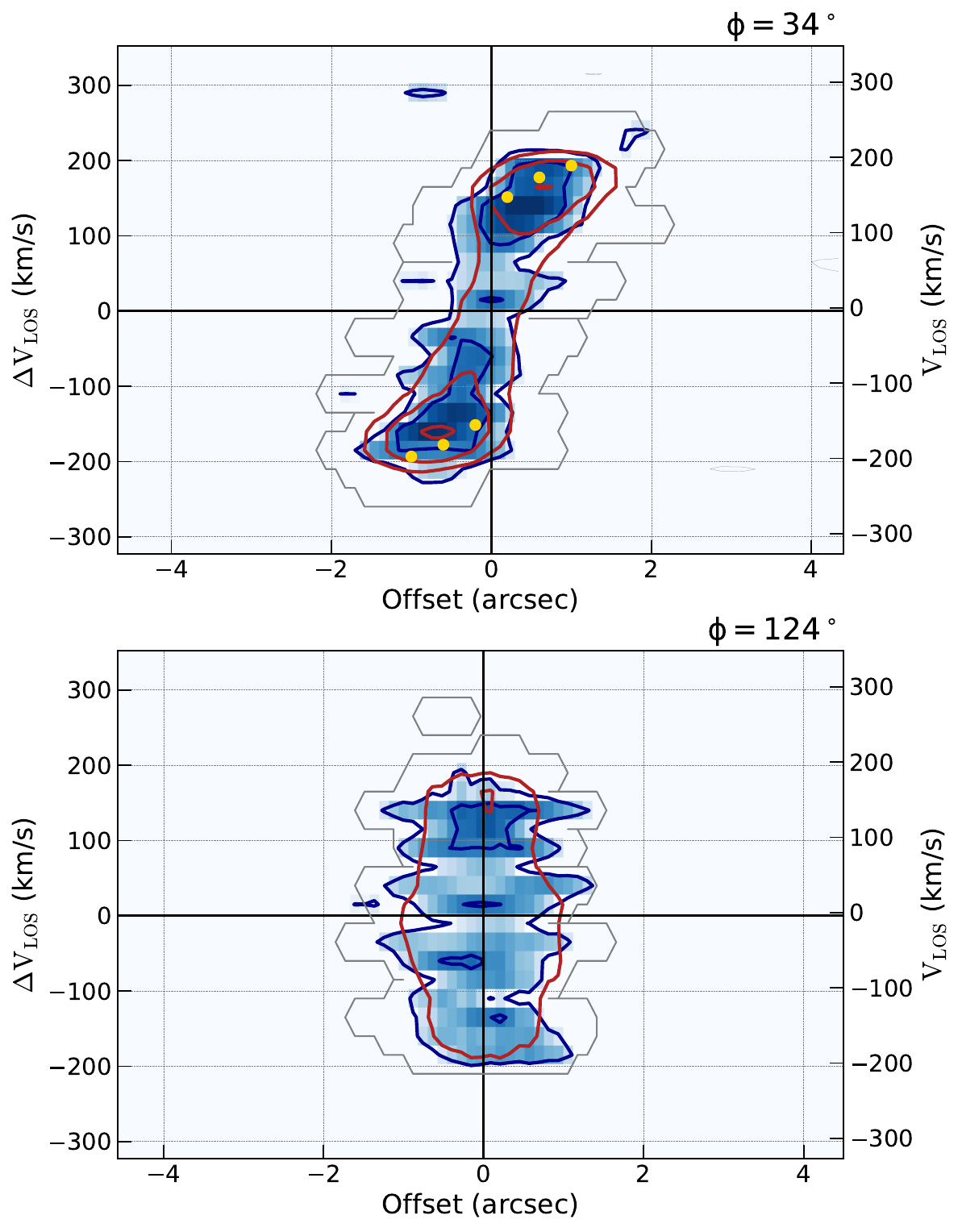}
\includegraphics[width=0.335\textwidth]{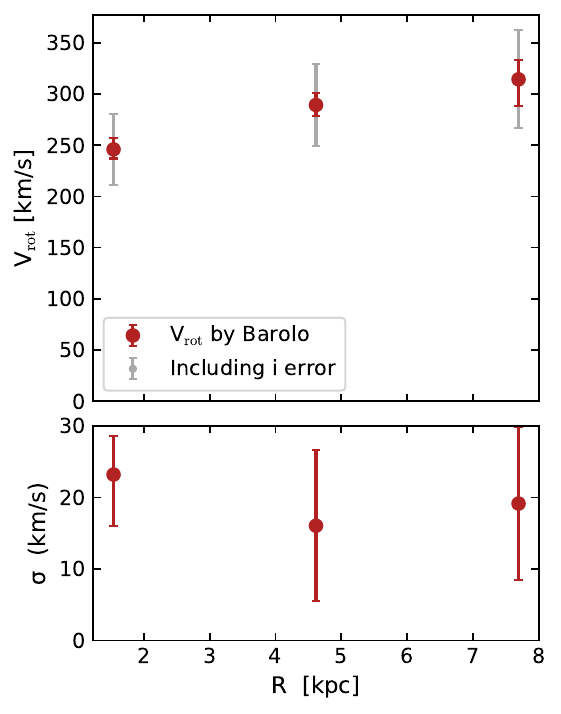}
  \includegraphics[width=0.60\textwidth]{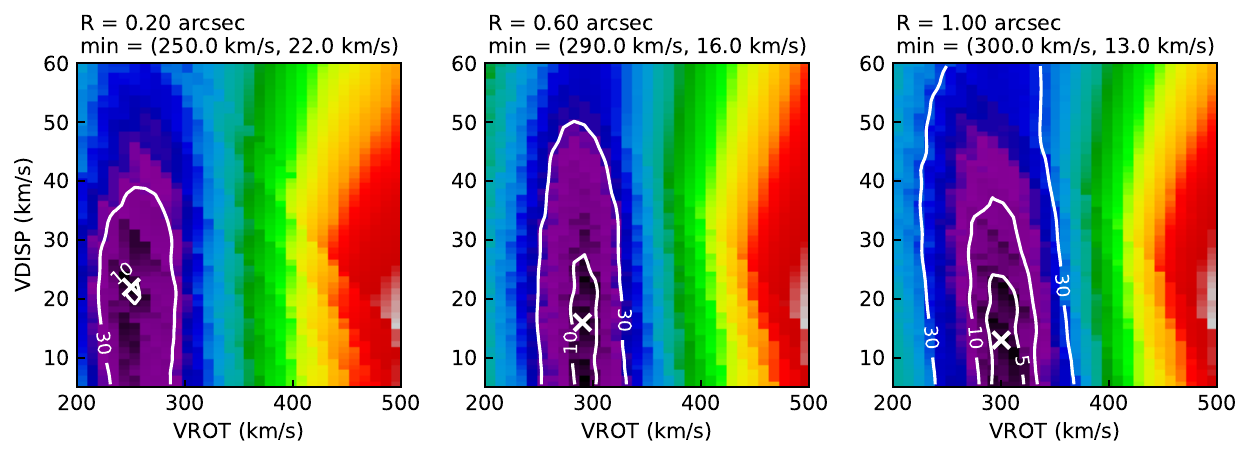}
    \caption{Data and best-fit model obtained with \barolo{}.
    \textit{Top left:} Intensity map (top) with S/N contours, calculated as in \citep{2001Verheijen,2014Lelli}, and velocity field (bottom) with isovelocity curves. The dashed line and the green cross indicate the kinematic $\phi$ and centre. 
    \textit{Top centre:} Position-velocity diagrams along the disc major (top) and minor (bottom) axes.  The observed emission is shown in blue, with the best-fit model (red contours) and ring LOS velocities (yellow points). The grey contour shows the applied mask. 
    \textit{Top right:} Rotation curve (top) and velocity dispersion (bottom) recovered by \barolo{} (red points). The grey errorbars include photometric disc inclination uncertainties.
    \textit{Bottom panels:} Parameter space for the rotation velocity and velocity dispersion per ring. White crosses indicate the best-fit minima.}
    \label{fig:barolo}
\end{figure*}

A precise measurement of the disc inclination is essential to robustly recover the CO rotation velocity. Since the angular resolution of ALMA data is not ideal for measuring the gas disc geometry, we use JWST imaging in the filter F322W2, which covers the rest-frame V to I bands and traces reasonably well the distribution of evolved stellar populations. 

We use the Python package \textsc{photoutils} on the F322W2 cutout image of the BW. 
Since previous works calculated the sky background using the full image, we refined the background calculation following a similar approach to \cite{2023Marasco}.  Figure~\ref{fig:jwst} shows each step of this procedure. 
A sky-dominated image is obtained by masking out an elliptical area encompassing the galaxy disc. This area is defined by fitting an elliptical aperture to the image, with the centre fixed on the brightest pixel. The pixel intensity distribution in the sky-dominated image is then fitted with a combination of a Gaussian plus Schechter model. The mean and standard deviation of the Gaussian provide the sky background, $I_\mathrm{sky}$, and its root-mean-square (rms) noise, $\delta_\mathrm{sky}$, respectively. The Schechter function accounts for the high-intensity tail of the distribution, arising from point-like sources (e.g. stars, unresolved galaxies) and resolved systems. After subtracted $I_\mathrm{sky}$ from the original image, we mask foreground stars and other galaxies using the \texttt{segmentation} tool. Then, a set of elliptical isophotes is fitted to the galaxy region using the \texttt{isophote} package. Each ellipses is spaced by 2 pixels and defined by the centre coordinates $x_0$ and $y_0$, the position angle $\phi$, and the ellipticity $\epsilon = 1 -b/a$, where $a$ and $b$ are the major and minor axis. In the fitting, we apply sigma clipping with threshold equal to 2 times the rms in each ellipse and perform 5 iterations. 

We determine the galaxy centre as the median $x_0$ and $y_0$. The inclination is calculated via Monte Carlo propagation of uncertainties on $\epsilon$, extracting $10^4$ random realisations of Gaussian distributions defined by the measured values of $\epsilon$ and their errors. Each random $\epsilon$ is converted into the corresponding inclination using $\cos i ^2= [(b/a)^2 - q_0^2]/(1-q_0^2)$ and assuming an intrinsic axis ratio $q_0=0.13$. This is a standard choice for disc galaxies in the local Universe, but recent works indicate it is appropriate also for discs at $z \lesssim 3$ \citep{2026vanAsselt}. 
We obtain median inclination of $i = 38 \pm 6~\deg$. 

For each ellipse with mean intensity $I(R)$, the associated uncertainty is obtained as $\delta_I (R) = \max( \delta, \delta_\mathrm{sky})$, where $\delta$ is the intensity rms in the ellipse \citep{2023Marasco}. We extract $I(R)$ in progressively larger ellipses until we obtain a signal-to-noise ratio (S/N) $I(R)/\delta_I(R) =1$  (Figure~\ref{fig:jwst}, third panel). We thus obtain that the maximum extent of the stellar distribution is $R_\mathrm{max} \approx 20$~kpc. The robustness of the photometry analysis is verified by checking that the cumulative light profile (curve of growth) flattens at the $I_{\rm sky}$ level.

We then compare the inclination derived from JWST imaging with that obtained from the ALMA total intensity map using \texttt{CANNUBI}\footnote{\url{https://www.filippofraternali.com/cannubi}}, a Markov Chain Monte Carlo python routine for estimating the geometry of discs \citep[for details, see][]{2023RomanOliveira}. We run \texttt{CANNUBI} fitting for $x_0$, $y_0$, $i$, $\phi$, and the recommended radial separation between the rings to be used for kinematic modelling. We find best-fit centre and inclination ($i = 38^{+8}_{-10} $) in perfect agreement with the isophote analysis.

\subsection{Gas kinematics from ALMA}\label{sec:gaskin}
We model the gas kinematics using \barolo \footnote{\url{https://editeodoro.github.io/Bbarolo/}}\citep{2015DiTeodoro}, a software designed to perform a tilted ring model fitting on emission line datacubes of rotating discs. \barolo{} is ideal to measure the gas kinematics when the angular resolution of the observation is low because, prior the fitting to data, the model is convolved with the instrumental resolution. This procedure allows to correct for the effect of beam smearing and instrumental broadening, which can bias the recovery of gas kinematics. \barolo{} models a galaxy disc as a series of rings, each described by geometrical ($x_0$, $y_0$, $i$, $\phi$) and kinematical parameters, which are the rotation velocity $V_\mathrm{rot}$, the velocity dispersion $\sigma$, the radial velocity in the disc plane $V_\mathrm{rad}$, and systemic velocity $V_\mathrm{sys}$. 
Given the low resolution of the data, we opted for reducing the number of free parameters and, therefore, we set geometrical parameters to the values resulting from isophote fitting, and assume $V_\mathrm{rad} = 0~\kms$. Based on \textsc{CANNUBI} results, we set the spacing between rings at \texttt{radsep}=0.4~\arcsec, which allows fitting the disc with three rings (\texttt{nradii=3}). We also fix the ring thickness to an arbitrary value lower than the angular resolution (\texttt{z0=0.1}~\arcsec), and assume a Gaussian distribution for the gas layer (\texttt{ltype=0.1}~\arcsec). 

Prior the fitting, \barolo{} performs a masking to the datacube. 
Here we use the masking option \texttt{mask=smooth\&search} and build the mask by smoothing the datacube by a \texttt{factor=1.5}, including all pixels with S/N above 3 (\texttt{snrcut=3}), and then growing the number of pixels in the mask adding the adjacent pixels with S/N above 2.5 (\texttt{growthcut=2.5}). We fit both the approaching and receding sides of the disc (\texttt{side=b}), giving more weight to the pixels along the major axis (\texttt{wfunc=2}), which contains most of the information on rotation. 
For the model normalisation, we and choose the option \texttt{norm=azim}, so that the model is normalised to the azimuthally averaged surface brightness prior the calculation of the residuals with the data. The residuals are calculated as the absolute value of the model minus the data using the option \texttt{ftype=2}, which gives similar weight to bright and faint pixels. Since adjacent channels in the datacubes are independent, we set \texttt{linear=0.42}. 

We do a preliminary run leaving $V_\mathrm{sys}$ as a free parameter, so that it is automatically found by \barolo{} algorithm. This procedure gives $V_\mathrm{sys} = -10~\kms$, which is assumed to obtained the fiducial model. After this preliminary run, we also adjust the kinematic position to $\phi = 34~\deg$, so that it is perpendicular to the isovelocity contour corresponding to $V_\mathrm{sys}$ in the velocity map and the emission in the position-velocity diagrams (PVDs) is as symmetric as possible with respect to the galaxy centre and systemic velocity (i.e. vertical lines at null offset and at $V_\mathrm{sys}$ in the PVDs in Figure~\ref{fig:barolo}) \footnote{We note that the datacube is reproduced equally well by a model with $\phi=24~\deg$ (from isophote fitting) and $V_\mathrm{rad}=-30~\kms$ (or $V_\mathrm{rad}=-50~\kms$, if we adopt the morphological $\phi=18~\deg$ from \textsc{CANNUBI}). Nevertheless, $V_\mathrm{rot}$ and $\sigma$ of these alternative models with radial motions are perfectly consistent with the fiducial values used here.}. 
Then, we obtain our fiducial model by running \barolo{} with only $V_\mathrm{rot}$ and $\sigma$ as free parameters, imposing a lower limit on $\sigma$ (\texttt{minvdisp=11}$~\kms$) based on the minimum recoverable value for data with channel width of $\approx 25 ~\kms$. The central panels in Figure~\ref{fig:barolo} show that our fiducial model (red contours) well reproduces the observations (in blue). The rotation curve is mildly increasing, reaching a maximum $V_\mathrm{rot} \approx 314~\kms$ at the outermost radius. We note that the uncertainties on $V_\mathrm{rot}$ calculated by \barolo{} in our fiducial fit are underestimated, as the uncertainties on $i$ ($\delta i$) on not taken into account. Therefore, for each ring, we add in quadrature $\delta V_\mathrm{rot} =  V_\mathrm{rot} \delta i \left(\tan i \right)^{-1}$ to the uncertainty on $V_\mathrm{rot}$ give by \barolo{}. The velocity dispersion is consistent (within the uncertainties) with being constant at $\sigma \approx 20~\kms$, but this is most likely because the data resolution is insufficient to resolve a radial gradient, rather than a genuinely constant $\sigma$. 
We use the \barolo{} task \texttt{spacepar} to inspect the parameter space and check whether the algorithm is converging to good minima. The bottom panels in Figure~\ref{fig:barolo} shows the $V_\mathrm{rot}$ versus $\sigma$ space. 
This indicates that $V_\mathrm{rot}$ is robustly recovered, while $\sigma$ is less constrained. Nevertheless, $\sigma \gtrsim 25~\kms$ are clearly disfavoured. 
Overall, these results confirm that the Big Wheel is a rotation-dominated discs with $V_\mathrm{rot}/\sigma \gtrsim 10$ \citep[W25;][]{Pensabene25}. 

\section{Refined SED fitting for the Big Wheel}
\label{appendix:SED}
\begin{figure}[h!]
    \centering
    \includegraphics[width=\columnwidth]{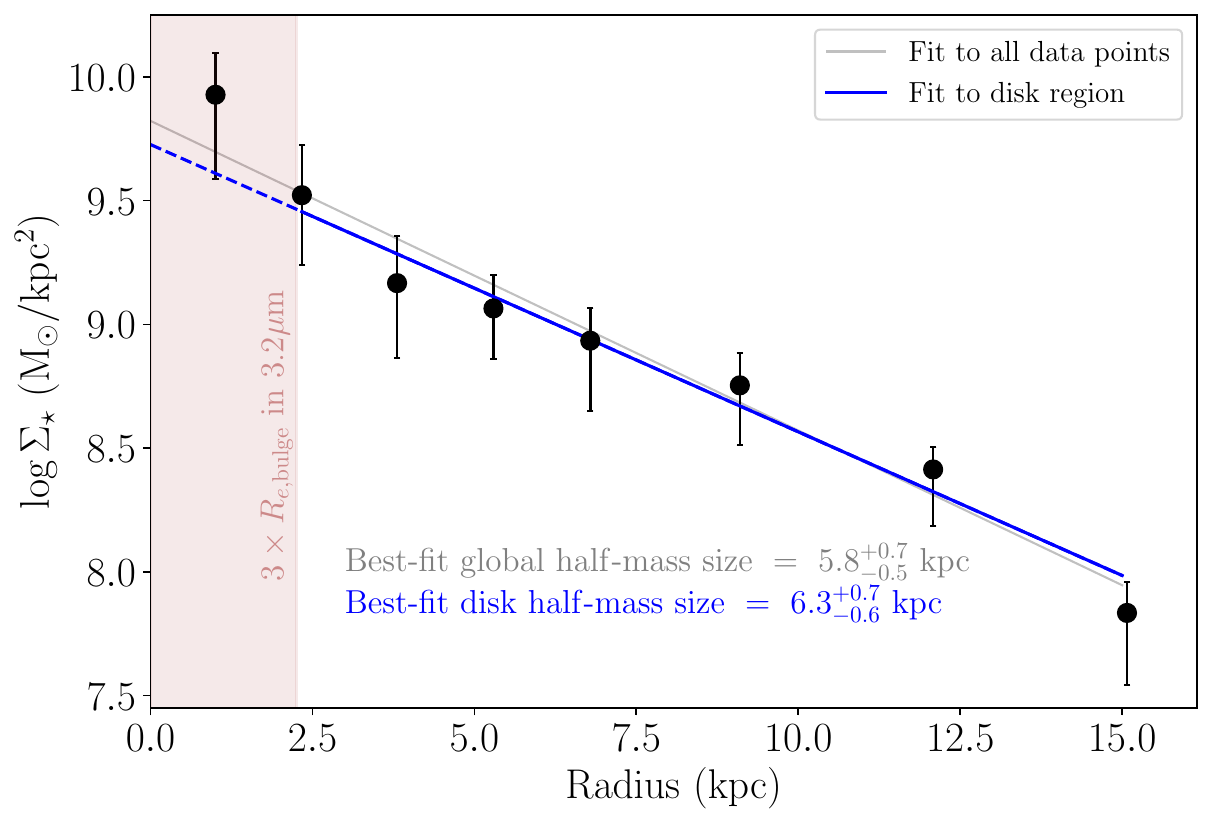} 
    \caption{Stellar mass density profile of the Big Wheel galaxy, obtained by performing SED fitting within finer radial bins than those used in W25, considering all available photometric bands.}
    \label{fig:BW_massprofile}
\end{figure}

To derive the stellar mass density profile of the Big Wheel galaxy while effectively disentangling the disc from the central regions — which could be affected by AGN emission and the presence of a bulge — we performed SED fitting using finer radial bins than those adopted in W25. Specifically, we considered bins of $1.5$~kpc up to $7.5$~kpc and $3$~kpc up to $16.5$~kpc, in contrast to the coarser binning of $2.25$~kpc up to $6.75$~kpc and $4.5$~kpc up to $15.75$~kpc. From this refined SED fitting, we recovered a stellar mass density profile that closely follows an exponential distribution, as shown in Figure~\ref{fig:BW_massprofile}. The solid grey line corresponds to the fit for the full dataset, while the blue line represents the fit excluding the leftmost point, with the dashed portion indicating the extrapolation). The resulting half-mass radius is equal to $r_{\rm half-mass}=6.3^{+0.7}_{-0.6}$~kpc.

\section{Posteriors of derived parameters}
\label{appendix:derpar}
In Figure~\ref{appendix:derpar}, we show the posterior distributions of the derived parameters for the fiducial rotation-curve decomposition presented in Figure~\ref{fig:vrotCO}. The values are summarised in Table~\ref{tab:posteriors}.
\begin{figure}[h!]
    \centering
    \includegraphics[width=\columnwidth]{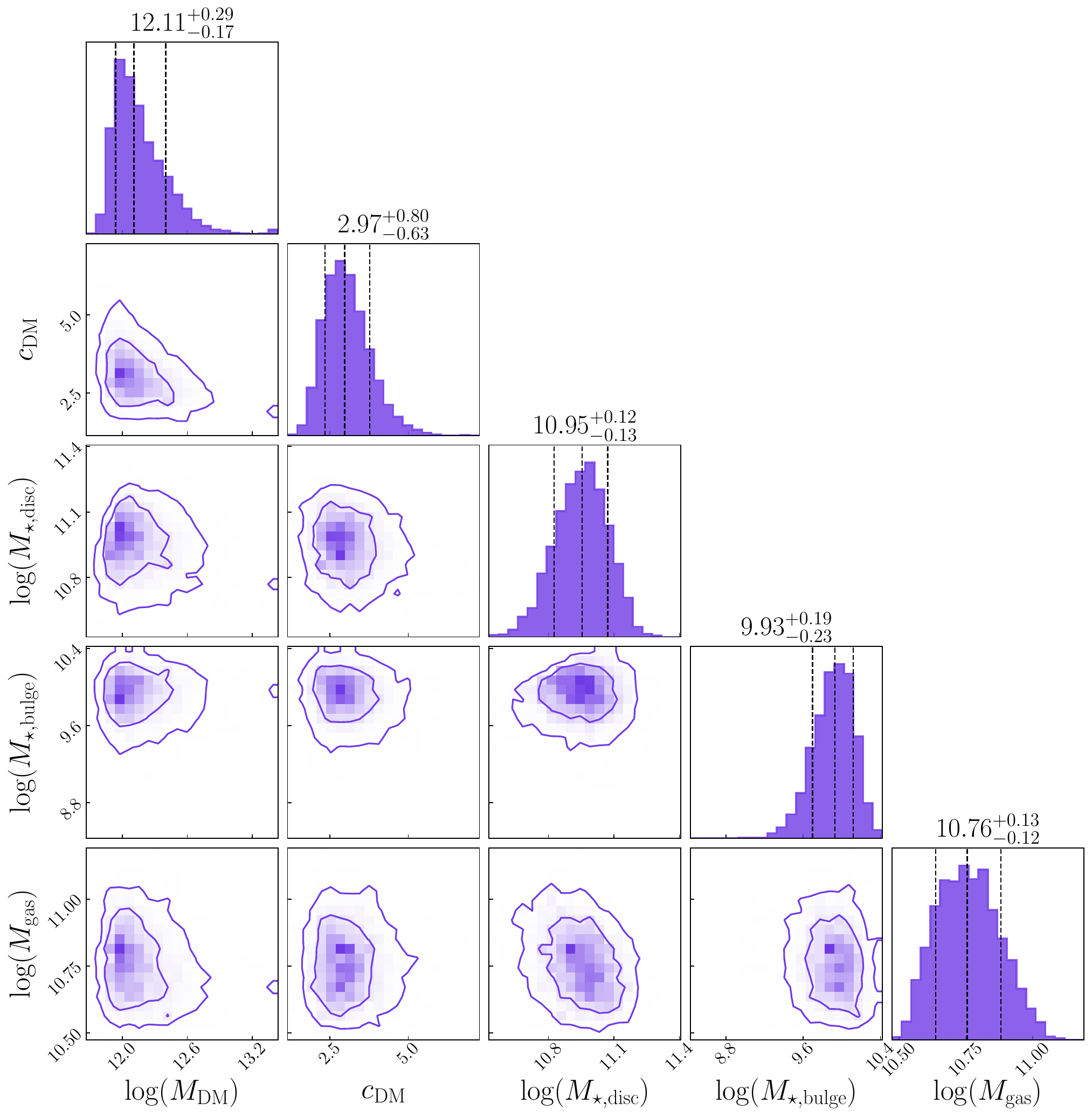} 
    \caption{Derived quantities of the Big Wheel galaxy, including, from left to right: the DM halo mass $\log (M_{\rm DM})$, its concentration $c_{\rm DM}$, the stellar disc mass $\log (M_{\star, \rm disc})$, the bulge mass $\log (M_{\rm bulge})$ and the gaseous mass $\log (M_{\rm gas})$.}
    \label{fig:vrotCO_fixedmstar}
\end{figure}

\section{Dynamical model for ADF22.A1 galaxy}
\label{appendix:adf22.a1}
\begin{figure}[h!]
    \centering
    \includegraphics[width=0.98\columnwidth]{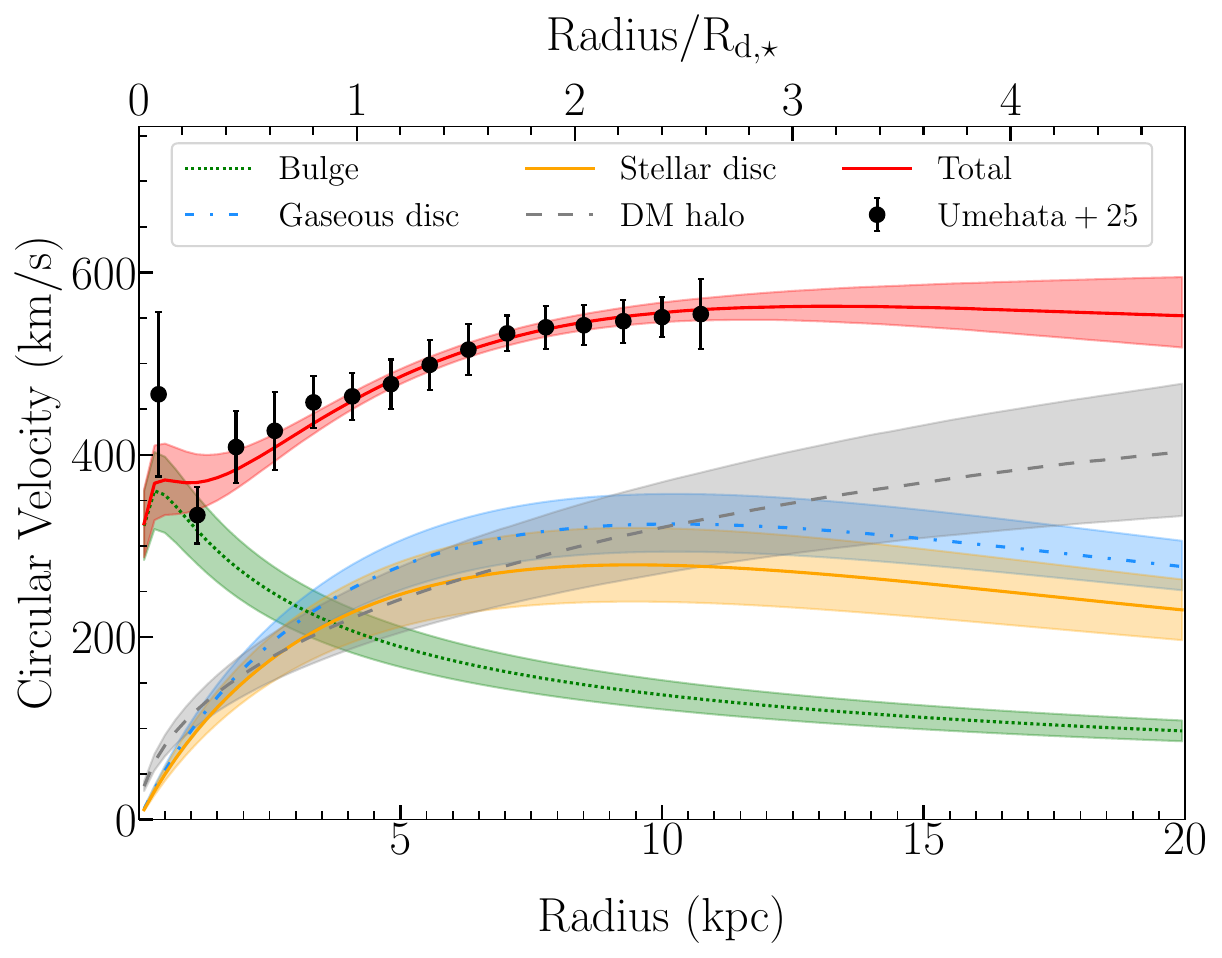} 
    \caption{Velocity curve of the ADF22.A1 galaxy derived from the best-fit parameters of the dynamical model performed with \texttt{dynesty}. The observational data points and the adopted priors are described in \cite{Umehata25}.}
    \label{fig:vrotCO_fixedmstar}
\end{figure}

With the aim of comparing the results obtained for the Big Wheel with similar objects at comparable redshifts, we applied our model to the ADF22.A1 galaxy, a giant and massive galaxy embedded in the SSA22 protocluster core, presented in \cite{Umehata25}. Similarly to the galaxy discovered in W25, ADF22.A1 has a stellar mass of $\log(\rm M_\star/\msun) =11.4^{+0.1}_{-0.2}$ and a physical diameter extending to at least $30$ kpc. The entire set of priors is built-up starting from the physical properties described in \cite{Umehata25}. In particular, the stellar and the gaseous masses, the effective radii for both discs inferred from the Sérsic profile fitting performed with \galfit{} and \barolo{} on different band images (F444W for the stellar disc and [CII] emission for the gaseous one), and the Sérsic index inferred for the bulge component. The presence of a bar possibly giving rise to the bulge has been suggested by \cite{Umehata25} based on the dust continuum and tentatively traced also in the gas kinematics (although \citealp{Rizzo26} did not find evidence of a bar in the photometry), and a MBH ranging among $5\times10^{10} \lesssim M_{\rm BH}/\msun \lesssim 4\times 10^{11}$ has been detected through X-ray luminosity. These two additional components have not been taken into consideration in the construction of the dynamical modelling, but their effects might be potentially limited to altering the distribution at small scales, the same region where the extracted rotational velocity is characterised by a large error bar.

In the posterior distributions, one crucial difference between the models of the Big Wheel and ADF22.A1 galaxies can be observed: the stellar mass corresponding to the best-fit is equal to $\log(\rm M_\star/\msun)=11.39^{+0.09}_{-0.11}$, in agreement with the value measured from the SED. The mass of its DM halo is equal to $\log(\mathrm{M}_{\rm DM}/\msun)=13^{+0.39}_{-0.36}$, leading to a SHRM equal to $\sim 0.025$ (see ~\ref{fig:SHMR_bw_adf22.a1} for a comparison with the Big Wheel). Regarding the concentration, we found $c=3.28^{+0.94}_{-0.77}$. We note that recently, \cite{Rizzo26} presented both a kinematic and a dynamical model for the ADF22.A1 galaxy. While our priors are based on the analysis by \cite{Umehata25}, their mass decomposition relies on a different kinematic derivation. Nevertheless, despite these differences in the underlying rotation curves and in the priors, our results remain consistent with their findings. Finally, for completeness and to allow a comparison with the Big Wheel galaxy, we extended the specific angular momentum analysis described in Section~\ref{sec:spin_jk} to this system. The results, based on the model from \cite{Umehata25}, are summarised in Table~\ref{tab:j_k_adf22.a1}. However, the refer to \cite{Rizzo26} for more precise $j_k$ estimates.

\newcolumntype{L}{>{\raggedright\arraybackslash}X}
\newcolumntype{Y}{>{\centering\arraybackslash}X}
\begin{table}[!t]
\def\arraystretch{1.3}
\begin{threeparttable}[b]
   \caption{As in Table~\ref{tab:j_k}, but for the ADF22.A1 galaxy. Note that the values reported in the table are based on the simplified model in \cite{Umehata25}.}
\label{tab:j_k_adf22.a1}
\centering
\begin{tabularx}{0.99\columnwidth}{l Y Y Y}
\toprule
\toprule
\multicolumn{4}{c}{Specific angular momentum results for ADF22.A1} \\
\midrule
\midrule
$j_k$ & Measured & Extrapolated (20 kpc) & Extrapolated (35 kpc)\tnote{a} \\
\midrule
$j_\star$ & $2804^{+15}_{-15}$ & $3992^{+42}_{-40}$ & $4430^{+65}_{-63}$ \\
$j_{\rm gas}$ & $2889^{+27}_{-27}$ & $4230^{+76}_{-78}$ & $4812^{+130}_{-129}$ \\
$j_{\rm bar}$ & $2818^{+14}_{-14}$ & $4034^{+39}_{-38}$ & $4499^{+63}_{-60}$ \\
\midrule
$j_{\rm DM}$ & - & \multicolumn{2}{c}{$4408^{+3852}_{-1979}$}\\
\bottomrule
\bottomrule
\end{tabularx}
\begin{tablenotes}
       \item [(a)] Convergence radius
     \end{tablenotes}
  \end{threeparttable}
\end{table}

\section{Baryon-only dynamical model}
\label{appendix:nodmmodel}
\begin{figure}[h!]
    \centering
    \includegraphics[width=0.98\columnwidth]{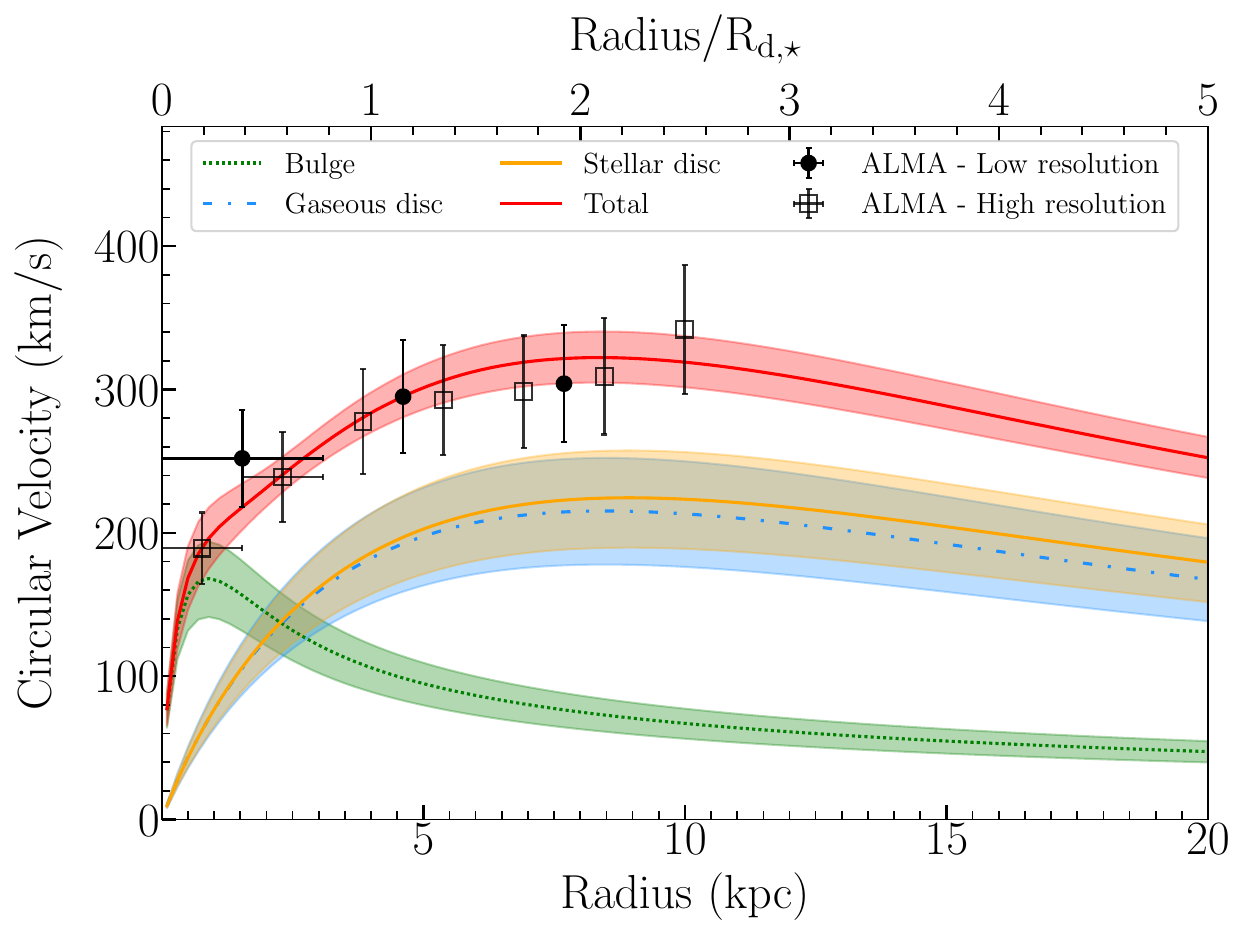} 
    \caption{Velocity curve of the Big Wheel as a function of the galactocentric radius, derived from the best-fit parameters of the dynamical model performed excluding the DM halo.}
    \label{fig:vrotCO_nodmmodel}
\end{figure}

Given that the posterior probability distribution for the baryon fraction $f_{\rm bar}$ (shown in Figure~\ref{fig:corner_plot}) smoothly extends to the highest allowed value and is only truncated by our prior ($f_{\rm bar} \leq f_{\rm bar, c}$), our results do not strictly demonstrate the necessity of a DM halo to explain the rotation curve of Big Wheel. Moreover, in every realisation of its dynamical modelling, the stellar mass is lower than the one measured by \cite{Galbiati25}. For this reason, we performed a dynamical test for an extreme scenario excluding the DM halo entirely. We adopted the same set of priors described in Section~\ref{sec:dyn_mod} for the masses of all baryonic components, while removing all parameters related to the NFW halo. The scale radii for the gaseous and the stellar discs, as well as for the stellar bulge, are fixed to the values obtained from the complete dynamical modelling (see in Figure~\ref{fig:corner_plot} and Table~\ref{tab:posteriors}). From this model, we find best-fit masses of $\log(\rm M_{\star}/\msun) = 11.14^{+0.11}_{-0.13}$ for the stellar mass, distributed between a stellar disc with $\log(\rm M_{\star, \rm disc}/\msun) = 11.10^{+0.12}_{-0.15}$ and a bulge with $\log(\rm M_{\star,\text{bulge}}/\msun)=10.02^{+0.12}_{-0.15}$, and $\log(\rm M_{\rm gas}/\msun)=10.91^{+0.14}_{-0.17}$ for the gaseous disc. The resulting rotation curve is shown in Figure~\ref{fig:vrotCO_nodmmodel}. Consistent with our previous findings, the stellar mass derived from the posterior distribution is lower than the value estimated through SED fitting, though it remains higher than that found in the previous test. This result is consistent with the viability of a baryon-only dynamical model. Specifically, the inclusion of a DM halo leads the best-fit model to favour a lower stellar mass.

We further performed the dynamical test without the NFW halo also for ADF22.A1. To compensate for the absence of the DM component, the posterior on the stellar mass yields a significantly higher value, $\log(\rm M_{\star}/\msun) = 11.61^{+0.06}_{-0.08}$. This estimate is twice as high as the value obtained from the SED fitting, suggesting that a scenario excluding the DM halo for ADF22.A1 leads to an unphysical overestimation of the stellar mass.

\end{appendix}

\end{document}